\documentclass[preprint,12pt]{elsarticle}
\usepackage{multicol}
\usepackage{amsfonts}
\usepackage{amssymb}
\usepackage{amsthm}
\usepackage{amsmath}
\usepackage{braket}
\usepackage{graphicx}
\usepackage{indentfirst}
\usepackage{mathrsfs}
\usepackage{graphics}
\usepackage{tikz}
\usetikzlibrary{decorations.pathreplacing,angles,quotes}
\usepackage{subcaption}
\usepackage{xcolor}
\usepackage{url}
\newcommand{\Tn}[0]{{Ti\textsubscript{76}Nb\textsubscript{22}Al\textsubscript{2}}}

\newcommand{\be}{\begin{eqnarray}}
\newcommand{\ee}{\end{eqnarray}}

\newcommand{\R}{{\mathbb R}}
\newcommand{\Rv}{{\mathbf R}}

\newcommand{\e}{{\mathbf e}}
\newcommand{\ub}{{\mathbf u}}
\newcommand{\vb}{{\mathbf v}}
\newcommand{\p}{{\mathbf p}}
\newcommand{\q}{{\mathbf q}}

\newcommand{\K}{{\mathbf K}}
\newcommand{\N}{{\mathbf N}}
\newcommand{\rb}{{\mathbf r}}
\newcommand{\bb}{{\mathbf b}}
\newcommand{\cb}{{\mathbf c}}
\newcommand{\db}{{\mathbf d}}
\newcommand{\s}{{\mathbf s}}
\newcommand{\ab}{{\mathbf a}}
\newcommand{\U}{{\mathbf U}}
\newcommand{\F}{{\mathbf F}}
\newcommand{\M}{{\mathbf M}}

\newcommand{\z}{{\mathbf z}}
\newcommand{\Q}{{\mathbf Q}}
\newcommand{\X}{{\mathbf X}}
\newcommand{\Z}{{\mathbf Z}}
\newcommand{\B}{{\mathbf B}}
\newcommand{\A}{{\mathbf A}}
\newcommand{\E}{{\mathbf E}}
\newcommand{\V}{{\mathbf V}}
\newcommand{\Pb}{{\mathbf P}}
\newcommand{\bzero}{{\mathbf 0}}
\newcommand{\Cb}{{\mathbf C}}

\newcommand{\n}{{\mathbf n}}
\newcommand{\x}{{\mathbf x}}
\newcommand{\y}{{\mathbf y}}
\newcommand{\cv}{{\mathbf c}}

\newcommand{\balpha}{{\boldsymbol\alpha}}
\newcommand{\blambda}{{\boldsymbol\lambda}}
\newcommand{\half}{\frac{1}{2}}
\newcommand{\av}{-\hspace{-.18in}\int}
\newcommand{\smallav}{-\hspace{-.15in}\int}
\newcommand{\ep}{\varepsilon}
\newcommand{\om}{\Omega}
\newcommand{\Div}{{\rm div}\,}

\newcommand{\supp}{{\rm supp}\,}

\newcommand{\rank}{{\rm rank}\,}
\newcommand{\dist}{{\rm dist}\,}

\newcommand{\1}{{\bf 1}}
\newcommand{\tr}{{\rm tr}\,}
\newcommand{\cof}{{\rm cof}\,}

\newcommand{\weakstar}{\stackrel{*}{\rightharpoonup}}
\newcommand{\esup}{\operatornamewithlimits{ess\, sup}}
\def \eps{\varepsilon}

\newcommand{\mint}[0]{\int_\Omega}

\newcommand{\vc}[1]{\mathbf{#1}}

\theoremstyle{definition}

\newtheorem{defn}{Definition}[section]

\newtheorem{thm}{Theorem}[section]
\newtheorem{rem}{Remark}[section]
\newtheorem{cor}[thm]{Corollary}
\newtheorem{lem}[thm]{Lemma}
\newtheorem{prop}[thm]{Proposition}

\usepackage{amssymb}
\usepackage{amsmath}

\journal{Journal of the Mechanics and Physics of Solids}

\graphicspath{%
    {converted_graphics/}
    {/}
}
\begin{document}

\begin{frontmatter}


 \title{Complex Martensitic Microstructures in \Tn\tnoteref{label1}}
 \author{John M. Ball}
 \ead{jb101@hw.ac.uk}
 \affiliation{organization={Heriot-Watt University and Maxwell Institute for Mathematical Sciences},
          city={Edinburgh},
            postcode={EH14 4AS},
            state={Scotland},
             country={U.K.}}


 \author{Francesco Della Porta}
 \affiliation{organization={Rho(Mi)},
country={Italy}}

\author{Tomonari Inamura} 
\ead{inamura.t.aa@m.titech.ac.jp}
\affiliation{organization={Laboratory for Materials and Structures,
Institute of Integrated Research,
Institute of Science Tokyo},
            addressline={J3-22, 4259 Nagatsutacho, Midori-ku}, 
            city={Yokohama},
            postcode={226-8501}, 
            country={Japan}}

\begin{abstract}
We analyze the unusual microstructure resulting from the cubic to orthorhombic phase transformation in the  shape-memory alloy \Tn. This microstructure seems to represent an approximate experimental realization of a deformation having a finite number of incompatible deformation gradients, which has been known to exist only as a theoretical possibility.

Because the middle eigenvalue $\gamma$ of the transformation strain   is very close to 1, the phase transformation  in this alloy  takes place via the nucleation of exact austenite-martensite interfaces rather than, as is usual for other alloys, by classical austenite-martensite interfaces separating the austenite from a finely-twinned laminate of martensite. According to the moving mask approximation \cite{DellaPorta2019} this leads to the question as to whether it is possible for a microstructure to be formed from the  set $H$ consisting of the twelve matrices $\V_i$ on the martensitic energy wells which are rank-one connected to the identity, even though these matrices are mutually incompatible, that is $\rank (\V_i-\V_k)=2$ for $i\neq k$.
 
We show that under the assumptions that $\gamma=1$ and that the transformation is volume preserving, such a microstructure (described by a gradient Young measure supported in $H$)  is possible with underlying macroscopic deformation that does not deform the grain boundary. This result is a consequence of $H$ forming a degenerate $T_{12}$ configuration in the sense of Kirchheim \cite{kirchheim03}. In fact there are such microstructures supported on just 8 of the 12 matrices in $H$. However there is no deformation satisfying the same boundary condition whose gradient  belongs to $H$ almost everywhere. An additional prediction, confirmed experimentally, is that the macroscopic deformation gradient equals the identity everywhere.

We also provide an analysis of the approximate compatibility of the $\V_i$. Although they are mutually incompatible, they are nearly compatible, with some pairs being much more nearly compatible than others, these pairs corresponding to interfaces that are more frequently observed. 
\end{abstract}




\end{frontmatter}
\centerline{\it Dedicated to Kaushik Bhattacharya on the occasion of his $60^{\rm th}$ birthday.}


\section{Introduction}
\begin{figure} [ht]
  \centering
 \includegraphics[width=4in,height=4in,keepaspectratio]{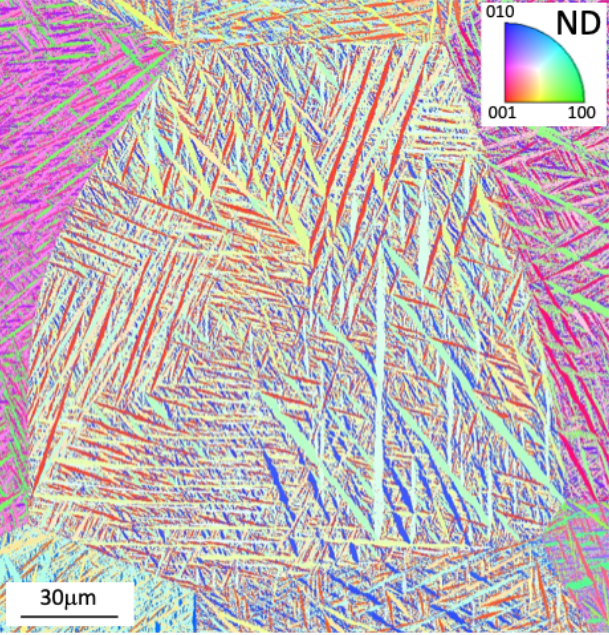}
  \caption{Inverse pole figure  map of the martensitic microstructure in \Tn, obtained by SEM-EBSD observation. The indices of the crystal planes exposed on the sample surface are visualized in accordance with the colour key in the top right-hand corner. }
  \label{fig1}
\end{figure}

The aim of this work is to describe and analyze the complex microstructure arising from the cubic to orthorhombic martensitic phase transformation in  the recently fabricated shape-memory alloy \Tn.  This microstructure (see Figure \ref{fig1}) seems to represent an approximate experimental realization of a deformation having a finite number of incompatible deformation gradients, which has been known to exist only as a theoretical possibility.
We frame our study in the context of continuum mechanics, using a  simplified model based on the energy-minimizing approach \cite{j32}.  For a single grain of {\Tn} occupying the bounded open domain $\om\subset\R^3$ in the reference configuration (taken to be the undistorted cubic austenite at the martensite start temperature $M_s$) we are interested in continuous deformations $\y:\om\to\R^3$ arising from the phase transformation. 
Because in this alloy the middle eigenvalue $\gamma=\lambda_2(\U)$ of the transformation strain 
$$\U=\left( \begin{array}{ccc} \alpha & 0 & 0 \\ 0 & \frac{\beta + \gamma}{2} &
\frac{\beta - \gamma}{2} \\ 0 &
\frac{\beta - \gamma}{2} & \frac{\beta + \gamma}{2}
\end{array} \right)$$
 is almost exactly one, the phase transformation takes place via the nucleation of exact austenite-martensite interfaces, rather than, as is usual for other alloys, by classical austenite-martensite interfaces separating the austenite from a finely-twinned laminate of martensite. 

Under the assumption that $\gamma=1$ the possible exact interfaces between the undistorted austenite  and the orthorhombic martensite phase are given by the rank-one connections between the identity matrix $\1$, belonging to
 the austenite  energy well $SO(3)$, and the martensite energy wells $SO(3)\U_i, i=1,\ldots,6$, where the $\U_i$ are the distinct matrices $\Q\U\Q^T$ for $\Q$ belonging to the cubic group $P^{24}$ of rotations.  As explained in detail in Section \ref{model},  there are exactly 12 such rank-one connections, two to each well $SO(3)\U_i$,
\begin{equation}
\label{variants0}
\1+\ab_{2i-1}\otimes\n_{2i-1}=\Rv_{2i-1}\U_i,\;\1+\ab_{2i}\otimes\n_{2i}=\Rv_{2i}\U_i,\;\;i=1,\ldots,6,
\end{equation}
with each $\Rv_i\in SO(3)$. We call each $\V_i:=\1+\ab_i\otimes\n_i$ a {\it $($habit-plane$)$ variant}. A key observation (see Proposition \ref{atensorn}) is that for this alloy $\rank (\V_i-\V_k)=2$ for $i\neq k$, so that the variants are mutually incompatible. 

According to
 the moving mask approximation of \cite{DellaPorta2019}, once the martensite is nucleated it undergoes no further deformation. Denoting by $H:=\{\V_1,\ldots,\V_{12}\}$ the set of variants, we are thus led to ask whether  it is  possible, despite the mutual incompatibility of the $\V_i$, for a microstructure to exist using just the gradients in $H$ and satisfying suitable boundary conditions on $\partial\om$. 

To formulate this problem precisely we describe microstructures using the language of {\it gradient Young measures}.  A gradient Young measure is a family of probability measures   $(\nu_{\x}\}_{\x\in\om}$ on $3\times 3$ matrices,  parametrized by $\x\in\om$, generated by a sequence of gradients in the sense described in Section \ref{TNsection}. The condition that the microstructure uses just gradients in $H$ is then the requirement that $\supp \nu_{\x}\subset H \text{ for a.e. }\x\in\om$,  so that 
$$\nu_{\x}=\sum_{i=1}^{12}\lambda_i(\x)\delta_{\V_i}$$
for local volume fractions $\lambda_i(\x)\geq 0$ with $\sum_{i=1}^{12}\lambda_i(\x)=1$. The underlying macroscopic deformation $\y$ then has deformation gradient $D\y(\x)=\bar\nu_{\x}$, where $\bar\nu_{\x}=\sum_{i=1}^{12}\lambda_i(\x)\V_i$ is the centre of mass of $\nu_{\x}$, and $D\y(\x)$ belongs for a.e. $\x\in\om$ to the {\it quasiconvexification} $H^{\rm qc}$ of $H$. The case of an {\it exact gradient} $D\y(\x)\in H$ for a.e. $\x\in\om$ corresponds to $\nu_{\x}=\delta_{\V_{i(\x)}}$ with $\V_{i(\x)}=D\y(\x)$.

The problem of studying gradient Young measures supported on sets of incompatible matrices, and corresponding exact gradients, has been studied by many authors. For the case of finitely many matrices $G=\{\A_1,\ldots,\A_N\}$ with $\rank(\A_i-\A_j)>1$ for $i\neq j$ a summary is given in \cite[Section 3.2]{j69}. Briefly, for $N=2$ it was shown in \cite[Propositions 1 and 2]{j32} that $\supp\nu_{\x}\subset G$ for a.e. $\x\in\om$ implies that $\nu_{\x}=\delta_{\A_i}$ for a.e. $\x$ and some $i$. The same result for the much more difficult case $N=3$ was proved by \v{S}ver\'{a}k \cite{Sverak93a,sverak93e}, making use of the result for exact gradients of Zhang \cite{zhang91} (for another proof see \cite{u6}). For  $N=4$ it was shown by Chlebik \& Kirchheim \cite{chlebik} that any exact gradient $D\y(\x)$ belonging to $G$ for a.e. $\x\in\om$ is affine. But the existence of $T_4$ configurations (see Section \ref{TNsection}) implies that  $\{\A_1,\A_2,\A_3,\A_4\}$ can be such that $\nu=\sum_{i=1}^4\lambda_i\delta_{\A_i}$ is a homogeneous (independent of $\x$) gradient Young measure with each $\lambda_i>0$.
For $N=5$ it was shown by Kirchheim \& Preiss (see \cite{kirchheim03}), using a version of the method of  {\it convex integration},  that $H=\{\A_1,\A_2,\A_3,\A_4,\A_5\}$ can be such that there is a nonconstant exact gradient with $D\y(\x)\in H$ for a.e. $\x\in\om$, when by the result of \cite{chlebik} all five values are taken in the neighbourhood of every point $\x$.
 Convex integration was introduced by Nash \cite{nash1954} and Kuiper \cite{kuiper} in the context of geometric analysis, generalized by Gromov \cite{gromov86}, and developed by M\"{u}ller \& \v{S}ver\'{a}k \cite{MullerSverak,MullerSverak99} and Dacorogna \& Marcellini \cite{dacorognamarcellini99} for use in the calculus of variations.   The case when $G$ is replaced by a finite number of incompatible compact sets of matrices is studied in \cite{j69}, while examples of non-affine smooth maps whose set $G$ of gradients has no rank-one connections are given in \cite{j35}.

For \Tn\,  the change of volume $\alpha\beta\gamma=\alpha\beta$ of the phase transformation is also extremely close to 1, and the assumption that $\alpha\beta=1$ leads to a more tractable theory. In particular, because (see Lemma \ref{sumrank})
$$\sum_{i=1}^{12}\ab_i\otimes\n_i=4(\alpha\beta-1)\1,$$
in the case $\alpha\beta=1$ we have that $\1$ is a convex combination of the $\V_i$. This implies that there is a {\it degenerate $T_{12}$ configuration} in $H$ (equivalently in $K=\{\ab_1\otimes\n_1,\ldots,\ab_{12}\otimes\n_{12}\}$) in the sense of Kirchheim \cite{kirchheim03}, which in turn implies see (Proposition \ref{LemmaRc}, \eqref{nu} and the related discussion) that there are gradient Young measures 
$$\nu_{\x}=\sum_{i=1}^{12}\lambda_i(\x)\delta_{\V_i}$$ 
 such that the corresponding macroscopic deformation $\y$ satisfies the boundary condition $\y(\x)=\x$ on $\partial\om$ (for why this is a reasonable boundary condition see Section \ref{expresults}).  For any such gradient Young measure there are explicit relations between the volume fractions $\lambda_i:=\smallav_{\om}\lambda_i(\x)\,d\x$, for example
$$\lambda_1+\lambda_4=\lambda_5+\lambda_7,\;\lambda_2+\lambda_3=\lambda_6+\lambda_8,$$
 (see Remark \ref{volumefracs} and Section \ref{6.2}). Furthermore (Proposition \ref{otherTN}) there are degenerate $T_8$ configurations in $H$, but no degenerate $T_N$ configurations for $N<8$. Whether or not $\alpha\beta=1$ there are no (nondegenerate) $T_N$ configurations in $H$ (Proposition \ref{noTN}), while if $\alpha\beta\neq 1$ there are (Lemma \ref{Pzero}) no degenerate $T_N$ configurations either.

If $\alpha\beta=1$ then, using a property of quasiregular maps and following an idea of Chlebik \& Kirchheim \cite{chlebik}, we are able to show (see Corollary \ref{rigidv}) that there is no map $\y$ with $D\y(\x)\in H$ for a.e. $\x\in\om$ satisfying the boundary condition $\y(\x)=\x$ for $\x\in\partial\om$, and that the only map satisfying the same boundary condition with $D\y(\x)\in H^{\rm qc}$ for a.e. $\x\in\om$ is $\y(\x)=\x$. The latter statement implies that if $\mu_{\x}=\sum_{i=1}^{12}\lambda_i(\x)\delta_{\V_i}$ is a zero-energy gradient Young measure with corresponding macroscopic deformation $\y$ that does not deform the grain boundary, then the macroscopic deformation gradient
 \begin{equation}
\label{averageF}
D\y(\x)=\sum_{1=1}^{12}\lambda_i(\x)\V_i=\1\text{ for a.e. }\x\in\om.
\end{equation}
This is confirmed experimentally (see Fig. \ref{fig5}). 
If $\alpha\beta\neq 1$ then it is natural to replace the boundary condition by the uniform dilatation $\y(\x)=(\alpha\beta)^{\frac{1}{3}}\x$ for $\x\in \partial\om$, but then Corollary \ref{rigidv} gives that there is no corresponding zero-energy microstructure.

Our analysis leaves open some apparently hard technical questions. First, we have been unable to determine the quasiconvex hull $H^{\rm qc}$ of $H$, knowledge of which would provide useful information on the boundary conditions consistent with a zero energy microstructure. A natural conjecture is that  $H^{\rm qc}$ consists in the case $\alpha\beta=1$  of the union of the 12 line segments joining $\1$ to the $\V_i$.  Second, we have been unable to decide whether it is true that every exact gradient with $D\y(\x)\in H$ for a.e. $\x\in\om$ is constant.

Although the $\V_i$ are mutually incompatible, they are almost compatible, and exact interfaces between variants are observed (so that the martensitic microstructure is not in fact stress-free). Under the assumption that the actual deformation gradients of the microstructure are close to those in $H$, this can be viewed as being consistent with the convex integration result of M\"{u}ller \& \v{S}ver\'{a}k that we give in Theorem \ref{MS THM}. In Section \ref{almost} we consider two different measures  of the almost compatibility of a pair of distinct variants $\V_i, \V_k$. The first is  $|\cof(\V_i-\V_k)|$, which we show is related to the Euclidean distance of $\V_i-\V_k$ from the cone $\Lambda:=\{\ab\otimes\n: \ab,\n\in \R^3\}$, and the second is the minimum angle of the rotations $\Q$ such that $\rank(\V_k-\Q\V_i)= 1$. According to either measure we find that certain pairs of variants, such as $\V_1, \V_k$ for $k=5,7,9,11$, are much more nearly compatible than others (with angles of rotation of about $0.5^\circ$). These pairs are the ones for which junctions are most frequently seen experimentally.

The plan for the paper is the following. In Section \ref{expresults} we summarize the experimental results. In Section \ref{model} we describe the cubic to orthorhombic phase transformation according to the nonlinear elasticity model of \cite{j32}, and  calculate the 12 rank-one connections $\V_i=\1+\ab_i\otimes\n_i$ from $\1$ to the martensitic energy wells $SO(3)\U_i$, showing that the $\V_i$ are mutually incompatible and establishing various properties for later use.  In particular, in the case $\alpha\beta=1$ we show (Lemma \ref{nullE}) that the null space of the map $\E:\R^{12}\to\R^{3\times 3}$ defined by $\E\x=\sum_{i=1}^{12}x_i\ab_i\otimes\n_i$ for $\x=(x_1,\ldots,x_{12})^T$ is 4-dimensional. We then study the almost compatibility of the $\V_i$, using Mallard's Law to give expressions for the angles of rotation between pairs of variants, the numerical values of the two measures of  almost compatibility being given in Table \ref{thetatable}.

In Section \ref{TNsection} we begin by recalling the definitions of the various semi-convex hulls of a compact subset $K$ of  $\R^{m\times n}$  and of gradient Young measures. Then we recall the definitions of  $T_N$ configurations and degenerate $T_N$ configurations, and give proofs of the key properties we use. Section \ref{analysis} contains our main results as described above, in which we make use of the technical apparatus of Section \ref{TNsection} and the detailed properties of the set $H$ of variants  established in Section \ref{model}. Finally, in Section \ref{Discussion} we discuss the implications of our analysis for the observed microstructure, and some of the issues involved in understanding its dynamic formation.\\

\section{Experimental results}
\label{expresults}
 Details of the alloy fabrication, lattice parameters, transformation temperatures and electron microscopy observations are summarized in the Appendix. 
   The variant notation follows that given in \eqref{variants0} and Section \ref{r1c}. In this section we use the term `variant $i$'  to refer also to regions in which the deformation gradient is close to $\1+\ab_i\otimes\n_i$.

\subsection{Orientation dispersion and boundary condition}
\label{dispersion}
\begin{figure} [htbp]
  \centering
 \includegraphics[width=5.5in,height=4.6in,keepaspectratio]{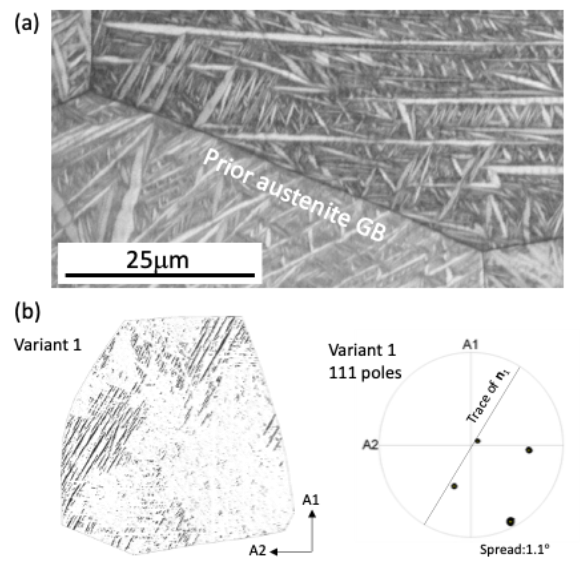}
  \caption{(a) IQ map of the grain boundary (GB) in the lower left-hand corner of Fig. 1. The prior austenite GBs remained smooth and fine mixtures of martensite are seen close to the GBs, with no obvious correlation between the microstructures on opposite sides, suggesting that the GBs had not deformed during the transformation. (b) Orientation dispersion of  variant 1 in the central grain in Fig. \ref{fig1} and its 111 pole figure with trace of $\n_1$. Martensitic plates belonging to the same variant have an orientation dispersion of only $1.1^\circ$.}
  \label{fig2}
\end{figure}
The specimen surface containing the central grain in Fig. \ref{fig1} is parallel to the (100 89 19) plane of the prior austenite, i.e., the normal to the observation plane is along [100 89 19] in the prior austenite cubic lattice.

 The individual martensite variants are observed monochromatically and the orientation dispersion for each variant is less than $1.1^\circ$. As explained in the introduction, the martensitic plates are not internally twinned because the middle eigenvalue of the transformation strain is very close to 1, thus permitting exact austenite-martensite interfaces. Previous studies in similar alloys  \cite{inamura2013} have shown that dislocations do not exist inside these martensitic plates. Although the EBSD measurements were performed at room temperature, the compatibility condition relevant to the formation of the martensitic microstructure corresponds to temperatures near $M_s$ at which austenite–martensite interfaces are present. We assume that the subsequent temperature dependence of the martensite lattice parameters does not significantly alter the established variant arrangement.

Figure 2(a) is an Image Quality (IQ) map of the grain boundary (GB) in the lower left-hand corner of Fig. 1. The trace of the prior austenite GB away from triple points is smooth. Many coarse martensitic plates  taper in thickness as they approach the GB, in the neighbourhood of which fine martensitic plates are densely concentrated. While it is possible for stress-free microstructures to deform grain boundaries (see the discussion of the Taylor set in \cite{u21}), the smoothness of the GB and the apparent lack of correlation between the microstructures on either side  suggest that it is undeformed, compatibility being maintained by a fine mixture of variants.  These considerations, and the fact that the volume change  $\det\U$ is very close to 1, make it reasonable to suppose that the deformation satisfies $\y(\x)=\x$ on the GB, as assumed in Section \ref{Discussion}.  

Figure 2(b) shows martensitic plates belonging to variant 1 extracted from the prior austenite grain at the centre of Fig. 1 and their 111-pole figure. The 111-poles are clustered with a very small spread of 1.1° or less. In other words, the orientation difference between martensitic plates belonging to variant 1 is negligible. The $\n_1$ trace (theoretical value, almost edge-on to the paper) drawn on the pole figure matches to the longitudinal direction of the martensitic plates of variant 1. This means that the martensitic plates are compatible with austenite and have approximately the same deformation gradient: $\Rv_1\U_1=\1+\ab_1\otimes \n_1$. Similar results were obtained for the other 11 variants. The deformation gradients in the martensitic microstructure can therefore be regarded as belonging approximately to the set consisting of the twelve matrices $\1+\ab_i\otimes\n_i$.

\subsection{Almost compatibility of the variants}
\label{almostc}
\begin{figure}[ht] 
  \centering
 \includegraphics[width=5.67in,height=3.6in,keepaspectratio]{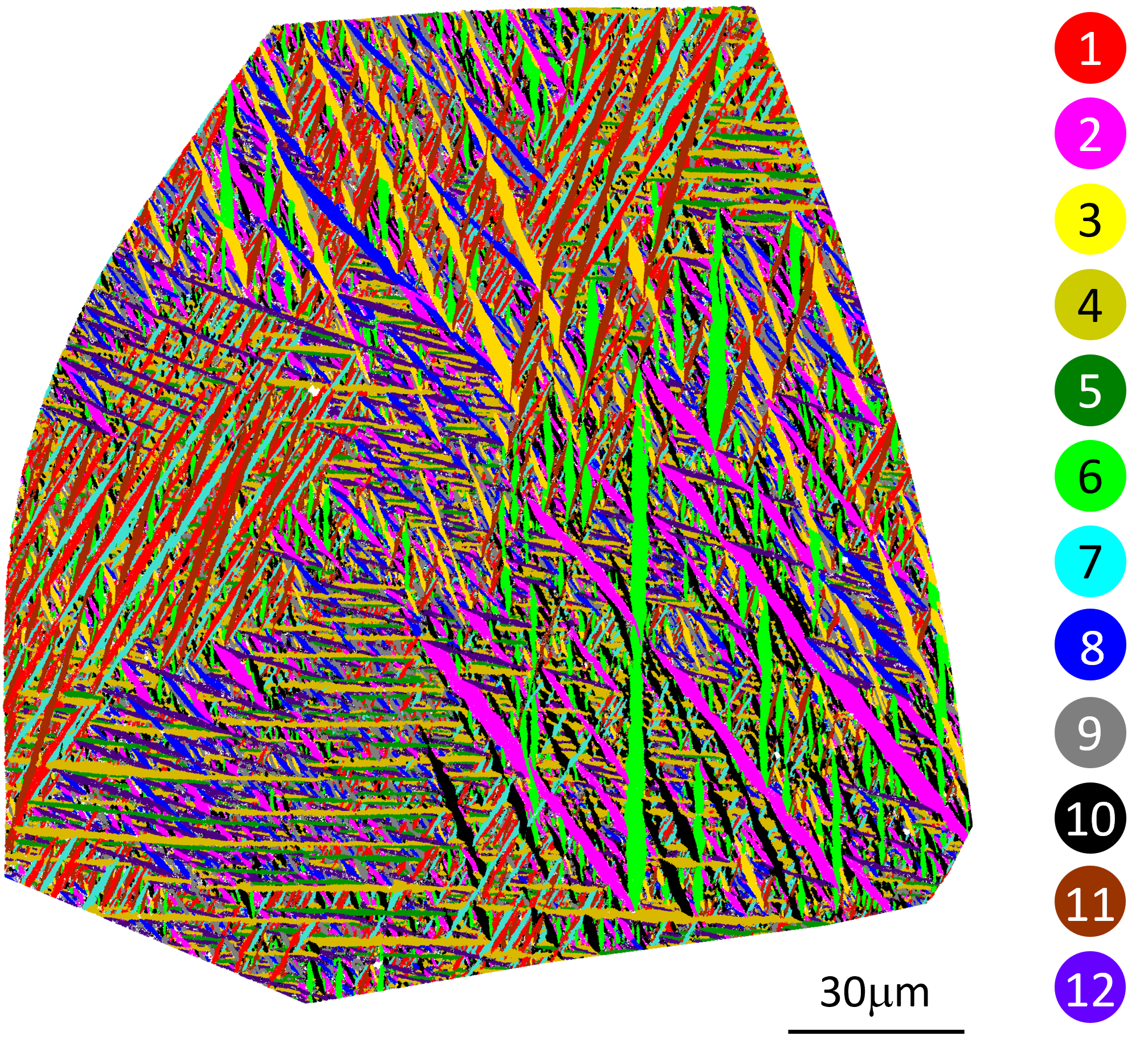}
  \caption{Variant map of the prior austenite grain in the centre of Fig. \ref{fig1}. Each variant was identified from the results of the orientation analysis by EBSD and colour-coded. Inter-variant connection of all combinations of the 12 variants occurred and is summarized in Fig. \ref{fig4}. The fractal character of the microstructure can be attributed to spatial partitioning by the martensitic plates. }
  \label{fig3}
\end{figure}

Figure \ref{fig3} shows the 12 colour-coded martensitic variants. The thickness of the plates differs from variant to variant because the angle between $\n_i$ and the plane of the observation differs for each variant. Variants 2 and 6, for example, appear thicker compared to variant 1 because $\n_2$ and $\n_6$ are nearly parallel to the specimen surface normal. 
It appears that as the transformation progresses the plates become obstacles to each other, making the formation of long plates more difficult. The fractal character of the microstructure in Fig. \ref{fig3} can be attributed to spatial partitioning by the martensitic plates in the spirit of the model \cite{p36,cesanahambly2022}  (see the discussion in Section \ref{stressed}).

As is shown in Proposition \ref{atensorn},   $\rank(\V_i-\V_k)=2$ for $i\neq k$, so that the 12 matrices $\V_i$  are mutually incompatible. However they are almost compatible, and Figure \ref{fig3}  clearly shows interconnections  with sharp interfaces between the  variants.  As a measure of the almost compatibility of the variants $\V_i$ and $\V_k$ we use the minimum $\theta_{ik}$ of the angles of rotation of the matrices $\Q\in SO(3)$  with
$$\rank(\Q(\1+\ab_i\otimes \n_i )-(\1+\ab_k\otimes \n_k ))=1. $$
 The angles $\theta_{ik}$ can be calculated for the measured lattice parameters (see Section \ref{free}) using the method in Section \ref{almost}, and are given in Table \ref{thetatable}.

\begin{figure}[t] 
  \centering
 \includegraphics[width=5.4in,height=2.7in,keepaspectratio]{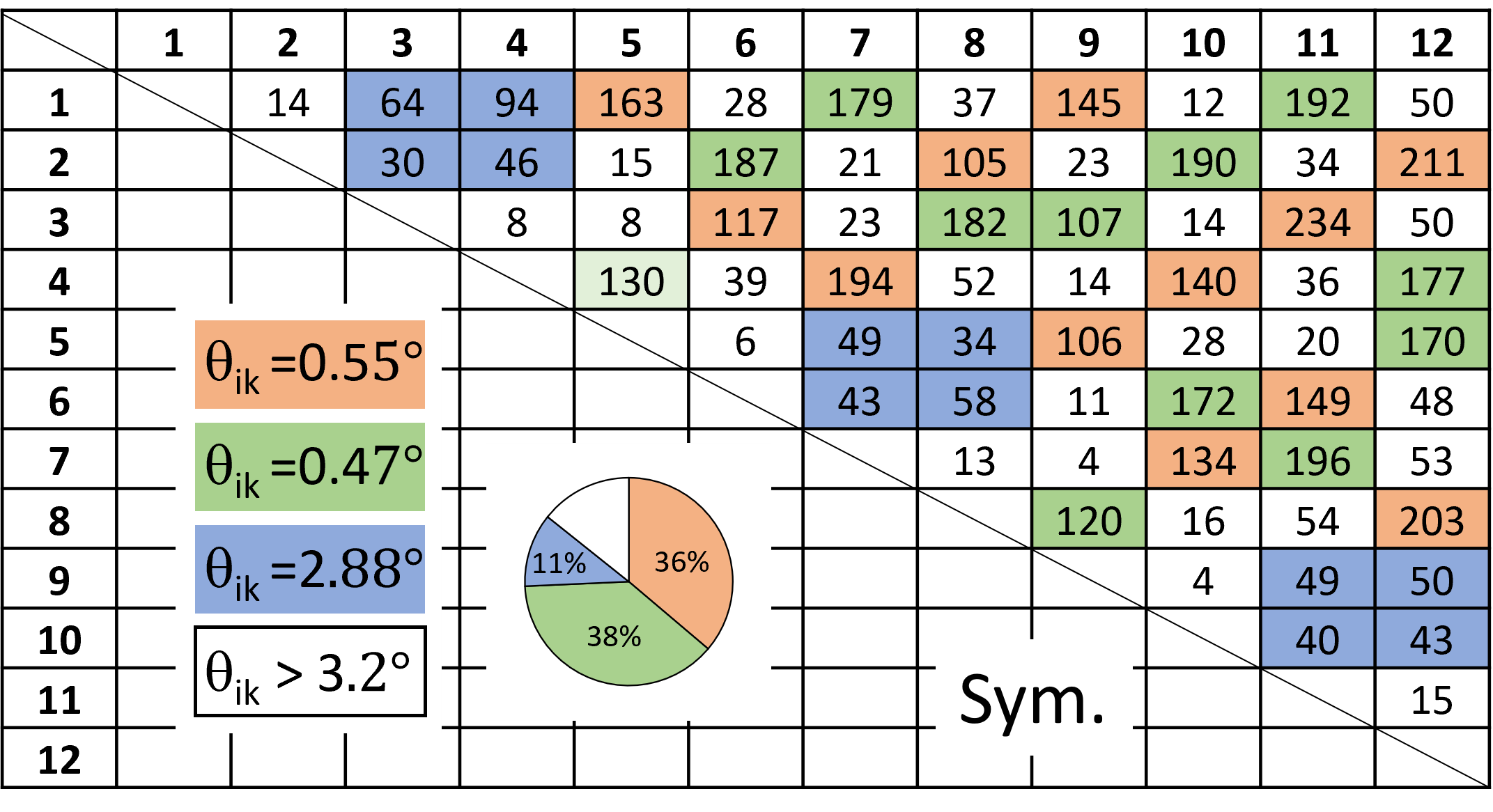}
  \caption{Frequency of inter-variant connection in the prior austenite grain at the centre of Fig. \ref{fig1}. All pairs of variants are incompatible. The total number of junctions detected was 5253, of which 36\% were with $\theta_{ik} ={0.55^\circ}$, 38\% with $\theta_{ik} ={0.47^\circ}$, 11\% with $\theta_{ik} ={2.88^\circ}$ and 15\% with $\theta_{ik} >3.2^\circ$. The two types of connections with the smallest incompatibility (green and orange) accounted for more than 70\% of the total.}
  \label{fig4}
\end{figure}

Figure \ref{fig4} summarizes the types of inter-variant connection that occurred in Fig. \ref{fig3} and their observed frequency, together with $\theta_{ik}$. Inter-variant connection of all combinations of the 12 variants occurred. As in the previous study [8], there was a clear trend in the frequency of occurrence; the two connections with the smallest incompatibility accounted for more than 70\% of the total. The total number of connections detected was 5253, of which 36\% were $\theta_{ik} ={0.55^\circ}$, 38\% were $\theta_{ik} ={0.47}^\circ$, 11\% were $\theta_{ik} ={2.88}^\circ$ and 15\% were $\theta_{ik} >3.2^\circ$. The interconnections with the smallest angles of rotation are also those for which another measure of almost compatibility considered in Section \ref{almost} is the smallest, the corresponding values of which are given in Table \ref{thetatable}.
\begin{figure}[ht] 
  \centering
 \includegraphics[width=5.67in,height=4in,keepaspectratio]{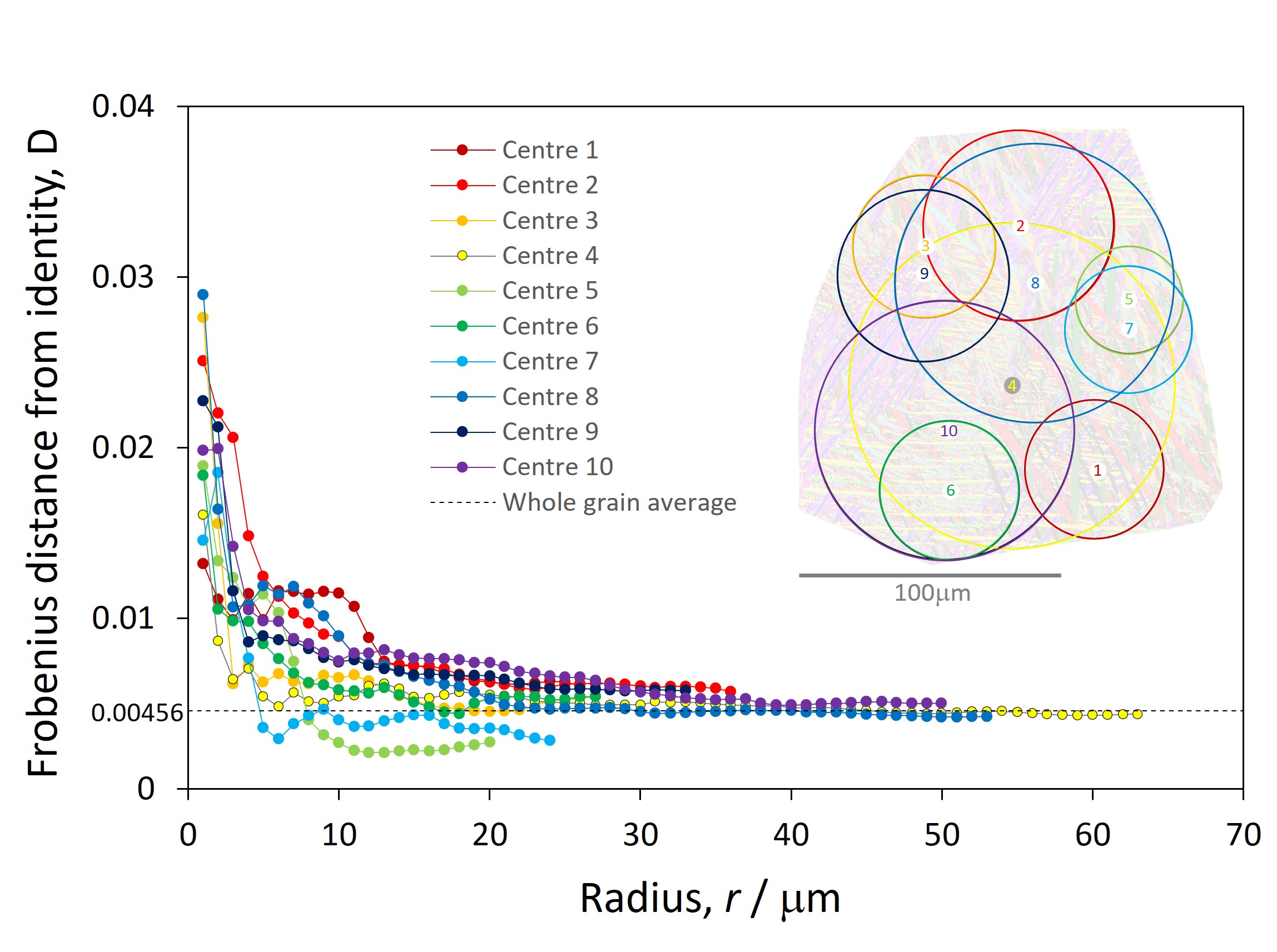}
  \caption{ Radius dependence of $D$ given by \eqref{defD} for the ten sampling centres shown in the inset. For each centre, $D$ was calculated from the area fractions of the twelve variants within a circular region of radius r. The circles in the inset indicate the maximum sampling radius for each centre. The horizontal dotted line indicates the whole-grain value, D=0.004556.}
  \label{fig5}
\end{figure}

A prediction of our theory (see Corollary \ref{rigidv} and \eqref{avgrad}) is that, under the assumption of volume preservation and undeformed grain boundaries, the macroscopic deformation gradient corresponding to the observed microstructure should equal the identity matrix $\1$.  The macroscopic deformation gradient at a point $\x$ corresponds to the average of the variants, taking into account their volume fractions, over a ball $B(\x,r)$ of small (but not too small) radius $r>0$. As explained in Section \ref{Discussion} we therefore expect that for any $\x_0$ on the surface of the sliced grain, and any not too small $r>0$, the average
 deformation gradient over the disk $B_2(\x_0,r)$ on the specimen surface should be close to $\1$, that is
\be
\label{expaver}
\sum_{i=1}^{12}\rho_i(\x_0,r)\V_i\sim\1,
\ee
where $\rho_i(\x_0,r)$ is the area fraction of $\V_i$ in $B_2(\x_0,r)$. To check this we evaluated
\be
\label{defD}
D=\left|\sum_{i=1}^{12}\rho_i(\x_0,r)\V_i-\1\right|
\ee
as a function of $r$ for 10 different points $\x_0$ on the specimen surface\footnote{This was done by counting the pixels corresponding to the different colours in the variant assignment of Fig. \ref{fig3}. A small number of pixels were blank and these were ignored when calculating the area fractions.}. Here $|\cdot|$ denotes the Frobenius norm (for the definition see Section  \ref{model}). The values of the $\V_i$ used are those corresponding to the deformation parameters given in Section \ref{free}. The results are shown in Figure \ref{fig4}, where it can be seen that, for each $\x_0$, $D$ rapidly becomes very small as $r$ increases, the value for the whole grain being $D=0.004556$. 

That this value is not exactly zero can be attributed to a combination of various factors such as  (i) the average is taken over a single two-dimensional disk rather than a three-dimensional ball (ii) the gradients are not exactly equal to the $\V_i$ (iii) the transformation is not exactly volume-preserving.

\section{The model}
\label{model}
\noindent{\it Notation}: $\R^{3\times 3}:=\{\text {real }3\times 3 \text{  matrices}\}$, $\R_+^{3\times 3}:=\{\A\in\R^{3\times 3}:\det\A>0\}$,  $SO(3):=\{\Rv\in\R^{3\times 3}:\Rv^T\Rv=\1,\; \det\Rv=1\}$ (the special orthogonal group of rotations) and $P^{24}:=\{\Rv\in SO(3):\Rv \text{ maps the cube $(-1,1)^3$ to itself}\}$ (the cubic group consisting of the 24 rotations with a single $\pm 1$ in each row and column). We denote by  $\A\cdot\B:=\tr(\A^T\B)$  the Frobenius inner product of $\A,\B\in\R^{3\times 3}$,  with corresponding norm $|\A|:=(\tr(\A^T\A))^\half$.
\subsection{Free energy and lattice parameters}
\label{free}
 We consider a single grain of {\Tn}   in which the undistorted cubic phase at the martensite start temperature $M_s$ occupies the bounded domain $\om\subset\R^3$.  According to the nonlinear elasticity model in \cite{j32}, for a deformation $\y:\om\to\R^3$ the bulk free energy is given by
$$I_\theta(\y)=\int_\om \psi(D\y(\x),\theta)\,d\x,$$
where $\theta>0$ is the temperature and $\psi(\A,\theta)$ is the free-energy density, defined for $\A\in \R_+^{3\times 3}$. By adding a function of $\theta$ to $\psi$ we may and shall assume that the minimum value of $\psi(\cdot,\theta)$ is zero for all $\theta$.  We suppose that $\psi$ is frame-indifferent and has cubic symmetry, so that
$$\psi(\Rv\A \Q,\theta)=\psi(\A,\theta)\text{ for all }\Rv\in SO(3), \Q\in P^{24}.$$ 
We assume an exchange of stability at $\theta=\theta_c$ so that the 
set 
\begin{align*}
K(\theta)&=\{\A\in \R^{3\times 3}_+: \A\text{ minimizes } \psi(\cdot,\theta)\}\\
&=\{\A\in\R^{3\times 3}_+:\psi(\A,\theta)=0\}
\end{align*}
 of energy-minimizing deformation gradients has the form 
$$K(\theta)=\left\{\begin{array}{ll}k(\theta)SO(3)&\theta>\theta_c\\k(\theta_c)SO(3)\cup\bigcup_{i=1}^NSO(3)\U_i(\theta_c)&\theta=\theta_c\\
\bigcup_{i=1}^NSO(3)\U_i(\theta)&\theta<\theta_c\end{array}\right.,$$
where $k(M_s)=1$, $\U(\theta)=\U(\theta)^T>0$ is the transformation strain, and the $\U_i(\theta), i=1,\ldots,N,$ are the distinct matrices $\Q\U(\theta)\Q^T$ for $\Q\in P^{24}$. 

 For a cubic to orthorhombic transformation $N=6$ and the $\U_i=\U_i(\theta)$ are given by
{\footnotesize
\begin{align*}
&\U_1 =\left( \begin{array}{ccc} \alpha & 0 & 0 \\ 0 & \frac{\beta + \gamma}{2} &
\frac{\beta - \gamma}{2} \\ 0 &
\frac{\beta - \gamma}{2} & \frac{\beta + \gamma}{2}
\end{array} \right),
&\U_2 =
\left( \begin{array}{ccc} \alpha & 0 & 0 \\ 0 & \frac{\beta + \gamma}{2} &
\frac{\gamma - \beta}{2} \\ 0 &
\frac{\gamma - \beta}{2} & \frac{\beta + \gamma}{2}
\end{array} \right),
 &\;\;\;\U_3 =
\left( \begin{array}{ccc} \frac{\beta + \gamma}{2} & 0 & \frac{\beta -
\gamma}{2} \\ 0 & \alpha & 0 \\
\frac{\beta - \gamma}{2} & 0 & \frac{\beta + \gamma}{2}
\end{array} \right), \\
&\U_4 =
\left( \begin{array}{ccc} \frac{\beta + \gamma}{2} & 0 & \frac{\gamma -
\beta}{2} \\ 0 & \alpha & 0 \\
\frac{\gamma - \beta}{2} & 0 & \frac{\beta + \gamma}{2}
\end{array} \right),
 &\U_5 =\left( \begin{array}{ccc} \frac{\beta + \gamma}{2} & \frac{\beta -
\gamma}{2}
& 0 \\ \frac{\beta - \gamma}{2} &
\frac{\beta + \gamma}{2} & 0 \\ 0 & 0 & \alpha
\end{array} \right),
&\;\;\;\U_6 =\left( \begin{array}{ccc} \frac{\beta + \gamma}{2} & \frac{\gamma -
\beta}{2}
& 0 \\ \frac{\gamma - \beta}{2} &
\frac{\beta + \gamma}{2} & 0 \\ 0 & 0 & \alpha
\end{array} \right),
\end{align*}
}\noindent
where the lattice parameters $\alpha>0,\beta>0,\gamma>0$ are the eigenvalues $\lambda_1<\lambda_2<\lambda_3$ of the $\U_i$, experimentally determined at $\theta=M_s=408$ K   as  $\alpha=0.9712, \beta=1.0296, \gamma=1.0003$.\footnote{See the Appendix for how these were determined. We choose the reference configuration as the undistorted austenite at $\theta=M_s$  both because  that is the temperature   at which the martensite is first nucleated and because the lattice parameters are reliably estimated there. In other work (such as \cite{j32}) the reference configuration is often taken to be the undistorted austenite at $\theta=\theta_c$, the temperature at which the austenite and martensite have the same free energy, so that $k(\theta_c)=1$.  With that choice $\gamma$ is still very close to $1$, the  lattice parameters at $\theta=\theta_c$ being estimated by  extrapolation to be $\alpha=0.9745, \beta=1.0272, \gamma=1.0002$. The EBSD measurements  were made at room temperature (approximately 300 K), and the microstructure  is assumed not to have been significantly affected by the reduction of temperature from $M_s$.
} Thus $\lambda_1=\alpha, \lambda_2=\gamma,\lambda_3=\beta$. Since $|\lambda_2-1|<4\times 10^{-4}$ {\it we assume from now on that} $0<\alpha<\gamma=1<\beta$. Also we note that $\det \U_i=\alpha\beta\gamma=1.0002$ is extremely close to $1$, so that often we will assume also that $\alpha\beta\gamma=\alpha\beta=1$. We expect that for temperatures $M_f<\theta<\theta_c$, where $M_f$ denotes the martensite finish temperature,  the energy well $k(\theta)SO(3)$ locally minimizes $\psi(\cdot,\theta)$ with $\psi(k(\theta)\1,\theta)>\psi(\U_i(\theta),\theta)$.

The eigenvectors of $\U_1$ are $\hat\e_1=\e_1$, corresponding to $\lambda_1=\alpha$, $\hat\e_2=\frac{\e_2-\e_3}{\sqrt 2}$ corresponding to $\lambda_2=1$ and $\hat\e_3=\frac{\e_2+\e_3}{\sqrt 2}$ corresponding to $\lambda_3=\beta$.

\subsection{Rank-one connections}
\label{r1c}
As described in Section \ref{expresults} we seek rank-one connections between $\1\in SO(3)$ and $\bigcup_{i=1}^6SO(3)\U_i$, that is vectors $\ab, \n\in\R^3, |\n|=1,$ such that
\be
\label{r1}
\1+\ab\otimes\n=\Rv\U_i
\ee
for some $\Rv\in SO(3)$.
By \cite[Proposition 4]{j32} there are exactly two such distinct rank-one connections  for each $i$. In the case $i=1$ they are given  (up to a change of sign) by
\begin{align}
\ab^\pm&=p\e_1\pm q(\e_2+\e_3)\label{aplus}\\
\n^\pm&=\rho^{-1}\left(-r\e_1\pm s(\e_2+\e_3)\right),\label{nplus}
\end{align}
where
\begin{align}&p=\beta\sqrt{\frac{(1-\alpha^2)(\beta-\alpha)}{\alpha+\beta}},\;q=\alpha\sqrt{\frac{(\beta^2-1)(\beta-\alpha)}{2(\alpha+\beta)}},\label{pq}\\
&r=\sqrt{1-\alpha^2},\; s=\sqrt{\frac{\beta^2-1}{2}},\; \rho=\sqrt{\beta^2-\alpha^2},\label{rs}
\end{align}
and we write $\ab_1=\ab^+, \n_1=\n^+$, $\ab_2=\ab^-,\n_2=\n^-$. In particular we have that $\1+\ab_1\otimes\n_1=\Rv_1\U_1$, $\1+\ab_2\otimes\n_2=\Rv_2\U_1$ for some $\Rv_1,\Rv_2\in SO(3)$. Given $\Q\in P^{24}$ we therefore have that 
$$\1+\Q\ab_1\otimes\Q\n_1=(\Q\Rv_1\Q^T)\Q\U_1\Q^T,$$
giving a rank-one connection from $\1$ to $SO(3) \Q\U_1\Q^T$. Considering first the four $\Q\in P^{24}$ with $\Q\U_1\Q^T=\U_1$ we find that 
\begin{align*}\Q\ab_1\otimes\Q\n_1&=\ab_1\otimes\n_1\text{ iff }\Q=\1\text{ or }-\1+(\e_2-\e_3)\otimes(\e_2-\e_3),\\
\Q\ab_1\otimes\Q\n_1&=\ab_2\otimes\n_2\text{ iff }\Q=-\1+2\e_1\otimes\e_1\text{ or }-\1+(\e_2+\e_3)\otimes(\e_2+\e_3).
\end{align*}
(Note that  the second $\Q$ changes the sign of both vectors, so that the tensor product is unchanged, and that three of these $\Q$ are $180^\circ$ rotations.)
Next, for each $i>1$ there exist  4 matrices $\Q_i\in P^{24}$ with $\Q_i\U_1\Q_i^T=\U_i$;  for definiteness we choose the $\Q_i$ to be the  $180^\circ$ rotations:
\begin{align}\nonumber
&\Q_2=-\1+2\e_2\otimes\e_2,&\\ 
&\label{Qis}\Q_3=-\1+(\e_1-\e_2)\otimes(\e_1-\e_2),& \Q_4=-\1+(\e_1+\e_2)\otimes(\e_1+\e_2),\\\nonumber&\Q_5=-\1+(\e_1-\e_3)\otimes(\e_1-\e_3),
&\Q_6=-\1+(\e_1+\e_3)\otimes(\e_1+\e_3).
\end{align}
(The nine $180^\circ$ rotations in $P^{24}$ are $-\1+2\e_j\otimes\e_j$ and $-\1+(\e_j\pm\e_k)\otimes(\e_j\pm\e_k),\,j\neq k$, from which it is easy to check that for $i>2$ there is just one such $180^\circ$ rotation, but for $i=2$ there are two, the other being $-\1+2\e_3\otimes\e_3$.)

Setting
$$\ab_{2i-1}=\Q_i\ab_1, \n_{2i-1}=\Q_i\n_1,\;\ab_{2i}=\Q_i\ab_2, \n_{2i}=\Q_i\n_2,$$
with $\Q_1=\1$, we get the two rank-one connections
  $$\1+\ab_{2i-1}\otimes\n_{2i-1}=\Rv_{2i-1}\U_i, \;\1+\ab_{2i}\otimes\n_{2i}=\Rv_{2i}\U_i$$
 from $\1$ to $SO(3)\U_i$, where $\Rv_{2i-1}=\Q_i\Rv_1\Q_i^T$, $\Rv_{2i}=\Q_i\Rv_2\Q_i^T$. Hence, for each $j=1,\ldots,12$, there are exactly two rotations $\Q\in P^{24}$ such that $\Q\ab_1\otimes\Q\n_{1}=\ab_{j}\otimes\n_{j}$, namely $$\Q_i,\;\Q_i(-\1+(\e_2-\e_3)\otimes(\e_2-\e_3)) \text{ if } j=2i-1 \text{ is odd, and}$$  $$\Q_i(-\1+2\e_1\otimes\e_1),\;\Q_i(-\1+(\e_2+\e_3)\otimes(\e_2+\e_3)) \text{ if }j=2i\text{ is even},$$   for the first  of which we have $\Q\ab_1=\ab_j, \Q\n_1=\n_j$ and for the second $\Q\ab_1=-\ab_j, \Q\n_1=-\n_j$. 

For later purposes we note that, since $\det(\1+\ab_i\otimes\n_i)=1+\ab_i\cdot\n_i=\alpha\beta$,
\begin{align*}|\U_1|^2&=\alpha^2+\beta^2+1=|\1+\ab_i\otimes\n_i|^2\\&=3+2\ab_i\cdot\n_i+|\ab_i|^2\\&=3+2(\alpha\beta-1)+|\ab_i|^2,
\end{align*} 
and so
\be
\label{asquared}
|\ab_i|^2=(\beta-\alpha)^2.
\ee

For $i=1,\ldots,12$ we write $\V_i:=\1+\ab_i\otimes\n_i$, $\Z_i=\ab_i\otimes\n_i$ and set
\be
\label{defK}
K:=\{\Z_1,\ldots,\Z_{12}\}=\{\ab_1\otimes\n_1,\ldots,\ab_{12}\otimes\n_{12}\},
\ee
and the corresponding set of deformation gradients by
\be
\label{defH}
H:=\{\V_1,\ldots,\V_{12}\},
\ee
so that $H=\1+K$.
\begin{prop}
\label{atensorn}
Let $0<\alpha<1<\beta$, and suppose that 
\be
\label{nonpar}
2\alpha^2+\beta^2\neq 3,\,2\alpha^{-2}+\beta^{-2}\neq 3.
\ee
Then there are no rank-one connections in $H$ or $K$, that is
 $$\rank(\ab_i\otimes\n_i-\ab_k\otimes\n_k)=2\text{ for }i\neq k.$$
\end{prop}
\begin{proof}
For there to be a rank-one connection we must have either $\ab_i\parallel\ab_k$ or $\n_i\parallel\n_k$ for some $i\neq k$. In the first case $\ab_i=\Q\ab_1, \ab_k=\Q'\ab_1$ for $\Q,\Q'\in P^{24}$ with $\Rv:=\Q'^T\Q\neq \1\text{ or }-\1+(\e_2-\e_3)\otimes(\e_2-\e_3)$ and $\Rv\ab_1=\lambda\ab_1$ for some $\lambda\neq 0$. Hence
\be
\label{pq1}
p\Rv\e_1+q(\Rv\e_2+\Rv\e_3)=\lambda(p\e_1+q(\e_2+\e_3)),
\ee
where $p, q$ are given by \eqref{pq}.
But since $2\alpha^{-2}+\beta^{-2}\neq 3$ we have $p\neq  q$. It then follows from  \eqref{pq1} that  $\Rv\e_1=\pm\e_1$. Indeed if, for example, $\Rv\e_1=\kappa\e_3$, $\kappa=\pm 1$, then $\Rv\e_2+\Rv\e_3=\kappa_1\e_1+\kappa_2\e_2$ with $\kappa_1=\pm 1, \kappa_2=\pm 1$, and \eqref{pq1} gives $\kappa p=\kappa_2 q=\lambda q$, which since $p$ and $q$ are positive contradicts $p\neq q$. If $\Rv \e_1=\e_1$ then \eqref{pq1} gives $\lambda=1$ and $\Rv\e_2+\Rv\e_3=\e_2+\e_3$, so that $\Rv=\1$. If $\Rv\e_1=-\e_1$ then $\lambda=-1$ and $\Rv\e_2+\Rv\e_3=-(\e_2+\e_3)$, giving $\Rv=-\1+(\e_2-\e_3)\otimes(\e_2-\e_3)$. Hence $\ab_i$ is not parallel to $\ab_k$ for $i\neq k$.

In the second case $\n_i\parallel\n_k$ we have similarly that $\Rv\n_1=\lambda\n_1$ for some $\Rv\in SO(3)$ with $\Rv\neq \1\text{ or }-\1+(\e_2-\e_3)\otimes(\e_2-\e_3)$, so that
$$-r\Rv\e_1+s(\Rv\e_2+\Rv\e_3)=\lambda(-r\e_1+s(\e_2+\e_3)),$$
where $r,s$ are given by \eqref{rs}. By assumption $r\neq s$, so that arguing as in the first case we get a contradiction.
\end{proof}
\begin{rem}
\label{roottwo}
In the case $\alpha\beta=1$ the second inequality in \eqref{nonpar} holds automatically, while the first reduces to $\beta\neq\sqrt{2}$.
\end{rem}
In the next result $[\ab,\bb,\cv]=(\ab\wedge\bb)\cdot\cv$ denotes the vector triple product.
\begin{prop}
\label{collinear}
Let $0<\alpha<1<\beta$ and let \eqref{nonpar} hold. \\
(i) If $\alpha^{-2}+2\beta^{-2}\neq 3$ then $[\ab_i,\ab_j,\ab_k]\neq 0$ for $i,j,k$ distinct.\\
(ii) If $\alpha^2+2\beta^2\neq 3$ then $[\n_i,\n_j,\n_k]\neq 0$ for $i,j,k$ distinct.
\end{prop}
\begin{proof}
We first claim that for  $\delta\neq 0,\pm 2$ and $\db:=\delta\e_1+\e_2+ \e_3$ we have $[\Q\db,\Rv\db,\db]\neq 0$ for $\Q,\Rv\in P^{24}$  provided no two of the vectors $\Q\db,\Rv\db,\db$ are parallel.  

Note that $\Q\db, \Rv\db, \db$ are all of the form $\pm (\delta\e_i +\kappa\e_j+\kappa'\e_k)$ with $i,j,k$ distinct and $\kappa=\pm 1$, $\kappa'=\pm 1$. There are three cases:\vspace{.05in}\\
Case 1. $\delta$ appears in the same slot for each vector, so that we may assume $\Q\db,\Rv\db$ are distinct vectors in the set $\{\delta\e_1+\e_2-\e_3, \delta\e_1-\e_2+\e_3, \delta\e_1-\e_2-\e_3\}$. Then $[\Q\db,\Rv\db,\db]=\pm 4\delta$. In particular $[\Q\db,\Rv\db,\db]\neq 0$ if $\delta=\pm 1$.  \vspace{.05in}\\
Case 2. $\delta\neq\pm 1$ appears in the same slot for $\Q\db,\Rv\db$ which is different from that for $\db$, so that we may assume that $\Q\db,\Rv\db$ are distinct vectors in the set $\{\e_1+\delta\e_2+\e_3, \e_1+\delta\e_2-\e_3, -\e_1+\delta\e_2+\e_3,-\e_1+\delta\e_2-\e_3\}$. Then $[\Q\db,\Rv\db,\db]$ takes one of the values $2(1-\delta^2), 2(\delta-1), 2\delta(1-\delta), 2\delta(1+\delta), 2(\delta+1), -2(1+\delta^2)$, and hence is nonzero since $\delta\neq 0$.\vspace{.05in}\\
Case 3. $\delta$ appears in three different slots, so that we may assume that $\Q\db=\e_1+\kappa_1\delta\e_2+\kappa_2\e_3$, $\Rv\db=\e_1+\kappa_3\e_2+\kappa_4\delta\e_3$ with each $\kappa_i=\pm 1$. Then  
$$[\Q\db,\Rv\db,\db]=\kappa_1\kappa_4\delta^3-(\kappa_1+\kappa_4+\kappa_2\kappa_3)\delta+\kappa_2+\kappa_3,$$
giving the possibilities $\delta^3-3\delta\pm 2=0$, $\delta^3+\delta\pm 2=0$, $\delta^3- \delta=0$ and $\delta^3+3\delta=0$, the real roots of which are $\delta=0, \pm 1, \pm 2$, completing the proof of the claim.

To prove (i), if $i,j,k$ are distinct then $\ab_i=\Rv_i\ab_1, \ab_j=\Rv_j\ab_1,\ab_k=\Rv_k\ab_1$ for $\Rv_i,\Rv_j,\Rv_k\in P^{24}$, so that $[\ab_i,\ab_j,\ab_k]=[\Q\ab_1,\Rv\ab_1,\ab_1]$, where $\Q=\Rv_k^T\Rv_i,\Rv=\Rv_k^T\Rv_j$ and we have used the invariance of triple products under rotations. By Proposition \ref{atensorn} no two of $\Q\ab_1,\Rv\ab_1,\ab_1$ are parallel. Since by \eqref{aplus}  $$[\Q\ab_1,\Rv\ab_1,\ab_1]=q^3[\Q\db,\Rv\db,\db]$$ with $\delta=p/q$, we have that $[\ab_i,\ab_j,\ab_k]\neq 0$ provided $p/q\neq 2$, that is, using \eqref{pq}, provided $\alpha^{-2}+2\beta^{-2}\neq 3$.

Similarly, using \eqref{nplus}, $[\n_i,\n_j,\n_k]\neq 0$ for $i,j,k$ distinct provided $r/s\neq 2$, that is provided  $\alpha^2+2\beta^2\neq 3$, proving (ii).
\end{proof}
\begin{rem}
If $\alpha\beta=1$ then the conditions in (i), (ii) hold iff $\beta\neq\sqrt{2}$, which is necessary and sufficient for \eqref{nonpar} also (see Remark \ref{roottwo}).
\end{rem}

\begin{lem}
\label{sumrank}
$$\sum_{i=1}^{12}\ab_i\otimes\n_i=4(\alpha\beta-1)\1.$$
\end{lem}
\begin{proof}
We note that by the discussion preceding Proposition \ref{atensorn}
\be
\label{defF}
\F:=\sum_{i=1}^{12}\ab_i\otimes\n_i=\half\sum_{\Q\in P^{24}}\Q\ab_1\otimes\Q\n_1,
\ee
from which it follows that  for any $\Rv\in P^{24}$
$$\Rv\F\Rv^T=\half\sum_{\Q\in P^{24}}\Rv\Q\ab_1\otimes\Rv\Q\n_1=\half\sum_{\Q\in P^{24}}\Q\ab_1\otimes\Q\n_1=\F.$$
Choosing first $\Rv=-\1+2\e_i\otimes\e_i$ we deduce that
$$2(\F\e_i\cdot\e_i)\e_i\otimes\e_i=\e_i\otimes \F^T\e_i+\F\e_i\otimes\e_i,$$
from which it follows by taking the scalar product with $\e_j, j\neq i,$ that $\F$ is diagonal. Since we can choose $\Rv$ so as to permute the diagonal entries, we get that $\F=\sigma\1$ for some $\sigma\in\R$. Since $\alpha\beta\gamma=\det(\1+\ab_i\otimes\n_i)=1+\ab_i\cdot\n_i$, $\tr\F=3\sigma=12(\alpha\beta-1)$, giving the result.
\end{proof}
It follows from Lemma \ref{sumrank} that in the case $\alpha\beta=1$ we have $$\sum_{i=1}^{12}\ab_i\otimes\n_i=\bzero,$$
so that $\bzero=\sum_{i=1}^{12}\frac{1}{12}\ab_i\otimes\n_i$ is a convex combination of the $\ab_i\otimes\n_i$, and thus belongs to the convex hull $K^{\rm c}$ of $K$ consisting of all such convex combinations (for an equivalent definition of $K^{\rm c}$ see Section \ref{TNsection}).

In the case $\alpha\beta=1$ it is useful to write the entries of the matrices $\ab_j\otimes\n_j$ explicitly. Setting $\ab_j\otimes\n_j=\half\left(\frac{1-\alpha^2}{1+\alpha^2}\right)\A^{(j)}$  for $j=1,\ldots,12$, the entries of the matrices  $\A^{(j)}$ are given in Table \ref{entries}, where $\sigma:=\sqrt{2}\alpha, \eta:=\frac{\sqrt{2}}{\alpha}$. We regard these entries as forming a $9\times 12$ matrix $\E$ representing the linear map from $\R^{12}$ to $\R^{3\times 3}\cong\R^9$ defined by $\E\x=\sum_{i=1}^{12}x_i\ab_i\otimes\n_i$ for $\x=(x_1,\ldots,x_{12})^T$.
\begin{table}[h]
\begin{tabular}{|c|c|c|c|c|c|c|c|c|c|c|c|c|}
\hline
\textit{j}      & 1         & 2        & 3         & 4         & 5        & 6         & 7         & 8         & 9        & 10        & 11        & 12        \\ \hline
$A^{(j)}_{11}$ & -2         & -2        & -2         & -2         & 1        & 1         & 1         & 1         & 1       & 1        & 1        & 1       \\ \hline
$A^{(j)}_{12}$ & $\eta$         & $-\eta$        & $-\eta$        & $\eta$        & $-\sigma$ & $\sigma$ & $-\sigma$  & $\sigma$ & $1$  & $1$    & $-1$   & $-1$    \\ \hline
$A^{(j)}_{13}$ & $\eta$ & $-\eta$ & $\eta$ & $-\eta$  & 1        & 1         & -1        & -1        & $-\sigma$  & $\sigma$    & $-\sigma$    & $\sigma$   \\ \hline
$A^{(j)}_{21}$ & $-\sigma$         & $\sigma$        & $\sigma$       & $-\sigma$        & $\eta$  & $-\eta$    & $\eta$   & $-\eta$    & $1$ & $1$ & $-1$  & $-1$ \\ \hline
$A^{(j)}_{22}$ & 1         & 1        & 1         & 1         & -2       & -2        & -2        & -2        & 1        & 1         & 1         & 1         \\ \hline
$A^{(j)}_{23}$ & $1$ & $1$ & $-1$  & $-1$ & $\eta$  & $-\eta$    & $-\eta$    & $\eta$   & $-\sigma$        & $\sigma$        & $\sigma$        & $-\sigma$        \\ \hline
$A^{(j)}_{31}$ & $-\sigma$    & $\sigma$  & $-\sigma$   & $\sigma$    & 1        & 1         & -1        & -1        & $\eta$ & $-\eta$ & $\eta$ & $-\eta$  \\ \hline
$A^{(j)}_{32}$ & $1$    & $1$  & $-1$    & $-1$   & $-\sigma$ & $\sigma$ & $\sigma$ & $-\sigma$  & $\eta$    & $-\eta$         &$ -\eta$        & $\eta$       \\ \hline
$A^{(j)}_{33}$ & 1        & 1       & 1        & 1        & 1        & 1         & 1         & 1         & -2        & -2      & -2         & -2         \\ \hline
\end{tabular}
\caption{The entries of the matrices $\A^{(j)}$, that form the $9\times 12$ matrix $\E$.}
\label{entries}
\end{table}
It will prove useful to calculate the nullspace $N(\E)$ of $\E$, that is  the set of vectors $\x$ such that
$\sum_{j=1}^{12}x_j\ab_j\otimes \n_j=\bzero$, which surprisingly depends only on the difference $\tau:=\eta-\sigma>0$.
\begin{lem}
\label{nullE}
The nullspace $N(\E)$ of $\E$ is 4-dimensional, and a basis is given by the four vectors $\bb^{(i)}=(b^{(i)}_j)\in\R^{12}$
\begin{align*}
\bb^{(1)}&=\left(2, 0, 0, 2, 2, 0, 2, 0, 1-\tau, 1-\tau, 1+\tau, 1+\tau\right)^T,\\
\bb^{(2)}&=(1-\tau,1-\tau, 1+\tau, 1+\tau, 2, 0, 0, 2, 2,0,0,2)^T,\\
\bb^{(3)}&=(2, 0, 2, 0, 1-\tau,1-\tau,1+\tau,1+\tau,2 ,0 ,2 ,0)^T,\\
\bb^{(4)}&=(1, 1, 1, 1, 1, 1, 1, 1, 1, 1, 1, 1)^T.
\end{align*}
\end{lem}
\begin{proof}
It can be checked explicitly that each $\bb^{(i)}\in N(\E)$. That the $\bb^{(i)}$  are linearly independent follows by noting that the determinant
\begin{align}
\label{det}
\det\left(\begin{array}{cccc}1-                                                                                                          \tau&2&2&1\\1-\tau&0&0&1\\1+\tau &0&2&1\\1+\tau&2&0&1\end{array}\right)=16\tau
\end{align}
formed from the last 4 rows is nonzero. This also implies that there are linear combinations of the $\bb^{(i)}$ of the form
\begin{align*}
\db_1&=( \cb_1,1, 0, 0, 0)^T,\\ \db_2&=( \cb_2,0, 1, 0, 0)^T,\\ \db_3&=(\cb_3,0, 0, 1, 0)^T,\\ \db_4&=(\cb_4,0, 0, 0, 1)^T,
\end{align*}
where the $\cb_i\in\R^8$. We claim that if  $\x=(x_1,x_2,\ldots,x_{12})^T\in N(\E)$ with $x_9=x_{10}=x_{11}=x_{12}=0$ then $\x=\bzero$. From this it follows that the $\bb^{(i)}$ span $N(\E)$, since if $\ub\in N(\E)$ then $\vb:=\sum_{i=9}^{12}u_i\db_i\in N(\E)$, so that $\x=\ub-\vb\in N(\E)$ satisfies $x_{9}=x_{10}=x_{11}=x_{12}=0$  and hence $\ub=\vb$.

To prove the claim, note that  $(\E\x)_5=(\E\x)_9=0$ gives that 
\begin{align}
x_1+x_{2}+x_{3}+x_{4}=x_5+x_6+x_7+x_8=0.
\end{align}
Also $(\E\x)_{2}-(\E\x)_{4}=(\E\x)_{3}-(\E\x)_7=(\E\x)_6-(\E\x)_8=0$ and $\eta+\sigma>0$ give
that
\begin{align*}
x_1+x_4+x_{6}+x_{8}&=x_2+x_3+x_{5}+x_{7}\\
x_2+x_{4}&=x_{1}+x_{3}\\
x_6+x_7&=x_5+x_8
\end{align*}
and hence 
\begin{align*}
x_6+x_7=x_5+x_8&=x_2+x_{4}=x_{1}+x_{3}=0,\\
x_6+x_8&=x_2+x_{3}.
\end{align*}
Then from $(\E\x)_{2}=0$ and $\eta>\sigma$ we get $x_6+x_8=0$ and from $(\E\x)_{3}=0$ that $x_5+x_6=x_7+x_8$, and hence $x_5=x_6=x_7=x_8=0$. Similarly, using $(\E\x)_6=0$ we deduce that $x_1=x_{2}=x_{3}=x_{4}=0$.
\end{proof}
\begin{rem}
\label{dimspan}
Since the dimension of $N(\E)$ is 4, by the rank-nullity theorem the range of $\E$ in $\R^{3\times 3}$ has dimension $12-4=8$. Therefore  the orthogonal complement of the linear span of the $\ab_j\otimes\n_j$ in $\R^{3\times 3}$ has dimension 1. But $\1\cdot \ab_j\otimes\n_j=\ab_j\cdot\n_j=0$ for all $j$, and hence  the only matrices that are orthogonal to every $\ab_j\otimes\n_j$ are multiples of the identity.
\end{rem}
\begin{lem}
\label{subsumrank}
Assume $\alpha\beta=1$,  $0<\alpha<1$.  

\noindent (i)  $\bzero$ is an interior point of $K^{\rm c}$ in the subspace $\{\A\in \R^{3\times 3}:\tr\A=0\}$.

\noindent (ii) If $\beta\neq\sqrt{2}$ then $\displaystyle\sum_{i\in P}\ab_i\otimes\n_i\neq\bzero$ for  any proper subset $P\subset \{1,2,\ldots,12\}$.

\end{lem}
\begin{proof}

Let $\M=(M_{ij})\in \R^{3\times 3}$  with $\tr\M=0$. 
To prove (i) it suffices to show by convex separation that   $\M\cdot\A^{(j)}\geq 0$ for all $j$ implies that $\M=\bzero$. First we note that $\sum_{j=1}^4\M\cdot\A^{(j)}\geq 0$, $\sum_{j=5}^8\M\cdot\A^{(j)}\geq 0$ and $\sum_{j=9}^{12}\M\cdot\A^{(j)}\geq 0$   imply respectively that $M_{22}+M_{33}\geq 2M_{11}$, $M_{11}+M_{33}\geq  2M_{22}$, $M_{11}+M_{22}\geq 2M_{33}$,  and hence, since $M_{11}+M_{22}+M_{33}=0$, we have that $M_{11}=M_{22}=M_{33}=0$. Then 
$\sum_{j=1}^2\M\cdot\A^{(j)}\geq 0$, $\sum_{j=3}^4\M\cdot\A^{(j)}\geq 0$ give $M_{23}+M_{32}=0$ and similarly we get $M_{12}+M_{21}=M_{13}+M_{31}=0$. Putting this back into $\M\cdot \A^{(1)}\geq 0$ and $\M\cdot\A^{(2)}\geq 0$ and noting that $\eta+\sigma>0$ we find that $M_{21}+M_{31}=0$, and similarly that $M_{32}+M_{12}=0$, $M_{13}+M_{23}=0$. Thus $\M=\bzero$.

To prove (ii) it suffices to show that if $\x\in N(\E)$ with each $x_i=0$ or 1 then  either $x_i=0$ for all $i$  or  $x_i=1$ for all $i$. By Lemma \ref{nullE} we have that $\x=\sum_{i=1}^4\alpha_i\bb^{(i)}$ for scalars $\alpha_i$. Then $x_{12}-x_{11}=2(\alpha_2-\alpha_3)\in \{0,\pm 1\}$,  $x_8-x_7=2(\alpha_2-\alpha_1)\in\{0,\pm 1\}$ and $x_{3}-x_{4}=2(\alpha_3-\alpha_1)\in\{0,\pm 1\}$, from which it follows easily that two of $\alpha_1,\alpha_2,\alpha_3$ are equal. Suppose without loss of generality that $\alpha_1=\alpha_2$. Then from $x_{9}-x_{11}=2\alpha_1(1-\tau)\in\{0,\pm 1\}$ and $x_{12}-x_{10}=2\alpha_1(1+\tau)\in\{0,\pm 1\}$ we deduce using $\tau\neq 1$ that $\alpha_1=\alpha_2=0$. But then $x_{10}=\alpha_4\in\{0,1\}$, $x_{11}=2\alpha_3+\alpha_4\in\{0,1\}$, $x_8=(1+\tau)\alpha_3+\alpha_4\in\{0,1\}$ and $\tau\neq 1$ imply that either $\alpha_3=\alpha_4=0$, or $\alpha_3=0, \alpha_4=1$ as required.
\end{proof}
\subsection{Almost compatibility}
\label{almost}
Although by Proposition \ref{atensorn} there are in general no rank-one connections in $H$,  pairs of elements of $H$ lying on different energy wells are {\it almost compatible} because the lattice parameters $\alpha,\beta$ are close to $1$.  

There are different ways to quantify the almost compatibility. We first note that given  $\X\in \R^{3\times 3}$ a necessary and sufficient condition that $\rank \X\leq 1$ is that $\cof \X=\bzero$.  Denoting the singular values of $\X$ by $v_1\geq v_2\geq v_3\geq 0$ we have that  \begin{align*}
|\cof \X|&=((v_2v_3)^2+(v_3v_1)^2+(v_1v_2)^2)^{\half}\\
&\geq \frac{\sqrt{3}}{2}(v_2^2+v_3^2)= \frac{\sqrt{3}}{2}\dist^2 (\X, \Lambda),
\end{align*}
where $\Lambda:=\{\Z\in\R^{3\times 3}:\rank \Z\leq 1\}$ denotes the closure of the  cone of rank-one matrices. Also, since $|\X|^2=v_1^2+v_2^2+v_3^2$,
 \begin{align*}
|\cof\X|^2&\leq 2v_1^2(v_2^2+v_3^2)\\
&\leq 2|\X|^2\dist^2(\X,\Lambda).
\end{align*}
Hence $|\cof\X|$ is a good measure of how far $\X$ is from a rank-one matrix. Setting $\X=\V_i-\V_k=\ab_i\otimes\n_i-\ab_k\otimes\n_k$ we have, using \eqref{asquared},
\begin{align*}
|\cof(\V_i-\V_k)|
&=|-(\ab_i\wedge\ab_k)\otimes(\n_i\wedge\n_k)|\\
&=(|\ab_i|^2|\ab_k|^2-(\ab_i\cdot\ab_k)^2)^{\half}(1-(\n_i\cdot\n_k)^2)^{\half}\\
&=(\beta-\alpha)^2\left(1-\left(\frac{\ab_i}{|\ab_i|}\cdot\frac{\ab_k}{|\ab_k|}\right)^2\right)^{\half}\left(1-\left(\n_i\cdot\n_k\right)^2\right)^{\half}.
\end{align*}
Taking without loss of generality $i=1$, the values of 
\begin{equation}
\label{deltak}
\delta_k:=|\cof(\V_1-\V_k)|
\end{equation}
 are given in Table \ref{thetatable}. Note that from  \eqref{aplus},\eqref{nplus} we have that $\delta_3=\delta_4$, $\delta_5=\delta_9,$ $\delta_7=\delta_{11},$ and $\delta_6=\delta_8=\delta_{10}=\delta_{12}$.

Another way \cite{j54, balandraudetal, teramoto}  to measure the almost compatibility is to note that if $i\neq k$ then 
\be
\label{orthconn}
\rank\left (\1+\ab_k\otimes\n_k-\Q(\1+\ab_i\otimes \n_i)\right)\leq 1
\ee
for some $\Q=\Q_{ik}\in SO(3)$. We have that  $\1+\ab_i\otimes \n_i\in SO(3)\U_I$ and $\1+\ab_k\otimes\n_k\in SO(3)\U_K$ for some $I,K$. If $I=K$ then, since $SO(3)\U_I$ has no rank-one connections, $\Q$ is the unique rotation such that $\1+\ab_k\otimes\n_k=\Q(\1+\ab_i\otimes \n_i)$. If $I\neq K$ then  there are exactly two such $\Q$ and \eqref{orthconn} expresses the fact that for cubic to orthorhombic transformations each distinct pair of energy wells $SO(3)\U_I, SO(3)\U_K$ is rank-one connected.  Then a measure of the almost compatibility is the angle of rotation of the unique such $\Q$ for $I=K$ and the minimum of the angles of rotation of the two such $\Q$ if $I\neq K$.

Without loss of generality we take $i=1$. Then for $j>1$
$$\U_j=(-\1+2\e\otimes\e)\U_1(-\1+2\e\otimes\e)$$ for the appropriate $180^\circ$ rotation from \eqref{Qis}. We use Mallard's law (see e.g. \cite[Proposition 2.2]{bhattacharya92}), according to which the rank-one connections between $\U_1$ and $SO(3)\U_j$  are given by
\begin{align}
\text{(Type I)}\hspace{.5in}&\nonumber\Q^I\U_j=\U_1+\ab^I\otimes\n^I,\\ &\Q^I=\left(-\1+2\frac{\U_1^{-1}\e}{|\U_1^{-1}\e|}\otimes\frac{\U_1^{-1}\e}{|\U_1^{-1}\e|}\right)\left(-\1+2\e\otimes\e\right),\label{typeone}\\ \text{(Type II)}\hspace{.5in}&\nonumber\Q^{II}\U_j=\U_1+\ab^{II}\otimes\n^{II},\\ &\Q^{II}=\left(-\1+2\frac{\U_1\e}{|\U_1\e|}\otimes\frac{\U_1\e}{|\U_1\e|}\right)\left(-\1+2\e\otimes\e\right).\label{typetwo}
\end{align}
 We first consider the case $k=2j-1$ for $j>1$.
Then we have that $\1+\ab_1\otimes\n_1=\Rv_1\U_1$, where $\Rv_1\in SO(3)$. Therefore the two rank-one connections from $\1+\ab_1\otimes\n_1$ to $SO(3)\U_j$ are given by 
\begin{align*}
\Rv_1\Q^I\U_j&=\1+\ab_1\otimes\n_1+\Rv_1\ab^I\otimes\n^I, \\\Rv_1\Q^{II}\U_j&=\1+\ab_1\otimes\n_1+\Rv_1\ab^{II}\otimes\n^{II}.
\end{align*}
But
\begin{align*}
\1+\ab_{2j-1}\otimes\n_{2j-1}&=(-\1+2\e\otimes\e)(\1+\ab_{1}\otimes\n_{1})(-\1+2\e\otimes\e),\\
&=(-\1+2\e\otimes\e)\Rv_1(-\1+2\e\otimes\e)\U_j,
\end{align*}
and therefore 
$$\rank(\1+\ab_{2j-1}\otimes\n_{2j-1}-\Q^{(2j-1)}(\1+\ab_{1}\otimes\n_{1}))=1,$$
where
\be
\label{Qjone}
\Q^{(2j-1)}=(-\1+2\e\otimes\e)\Rv_1(-1+2\z\otimes\z)\Rv_1^T,
\ee
and $\z=\frac{\U_1^{-1}\e}{|\U_1^{-1}\e|}$ or $\z=\frac{\U_1\e}{|\U_1\e|}$. By the Rodrigues formula (see e.g. \cite{MarsdenRatiu}) the angle of rotation $\theta$ of any $\Q\in SO(3)$ is given by
$$\cos\theta=\half(\tr\Q-1).$$
Hence the angle of rotation of $\Q^{(2j-1)}$ is given by 
\be
\label{angle}
\cos\theta=2(\e\cdot\Rv_1\z)^2-1,
\ee
so that
\be
\label{anglehalf}
\theta=2\cos^{-1}(\e\cdot\Rv_1\z).
\ee
We write $\theta=\theta_1$ for $\z=\frac{\U_1^{-1}\e}{|\U_1^{-1}\e|}$  (Type I)   and $\theta=\theta_2$ for  $\z=\frac{\U_1\e}{|\U_1\e|}$ (Type II). 

Now suppose $k=2j$ is even. The case $j=1$ is special because $\1+\ab_1\otimes\n_1=\Rv_1\U_1$ and $\1+\ab_2\otimes\n_2=\Rv_2\U_1$ lie on the same energy well and so $\1+\ab_2\otimes\n_2=\Rv_2\Rv_1^T(\1+\ab_1\otimes\n_1)$, giving
\begin{align*}
&\half(\tr\Rv_2\Rv_1^T-1)=\cos\theta=f(\alpha,\beta),\text{ where}\\&f(\alpha,\beta):=\half\left(\alpha\beta+\frac{1}{\alpha\beta}\right)-\frac{1}{2\alpha\beta}\left(\frac{\beta-\alpha}{\alpha+\beta}\right)^2(1+\alpha\beta)^2\simeq 0.9983,
\end{align*}
and hence $\theta\simeq 3.34^\circ$.

If $j>1$ then 
\begin{align*}
\1+\ab_{2j}\otimes\n_{2j}&=(-\1+2\e\otimes\e)(\1+\ab_{2}\otimes\n_{2})(-\1+2\e\otimes\e),\\
&=(-\1+2\e\otimes\e)\Rv_2 (-\1+2\e\otimes\e)\U_j,
\end{align*}
and therefore 
$$\rank(\1+\ab_{2j}\otimes\n_{2j}-\Q^{(2j)}(\1+\ab_1\otimes\n_1))=1,$$
where
\be
\label{Qjtwo}
\Q^{(2j)}=(-\1+2\e\otimes \e)\Rv_2(-\1+2\z\otimes\z)\Rv_1^T,
\ee
and $\z=\frac{\U_1^{-1}\e}{|\U_1^{-1}\e|}$ or $\z=\frac{\U_1\e}{|\U_1\e|}$, with angle of rotation $\theta$ given by
\be
\label{angleeven}
\cos\theta=\half(\tr(\Rv_2\Rv_1^T)-1)-\Rv_1^T\e\cdot\Rv_2^T\e-\Rv_1\z\cdot\Rv_2\z+2(\Rv_1\z\cdot\e)(\Rv_2\z\cdot\e).
\ee
The calculated values in degrees for different $k>2$ (together with the corresponding $\e$) are given in Table \ref{thetatable}.
\begin{table}[h]
\centering
\begin{tabular}{|c|c|c|c|c|c|}
\hline
$k$&$\e$&$\theta_1$(deg)&$\theta_2$(deg)& $\min\,(\theta_1,\theta_2)$&$\delta_k$\\ \hline 
 3&$\e_2$&2.88&2.91&2.88&0.0026\\ \hline
4&$\e_2$&2.91&2.88&2.88&0.0026\\ \hline
5&$\frac{1}{\sqrt{2}}(\e_1-\e_2)$&0.55&5.67&0.55& 0.00096\\ \hline
6&$\frac{1}{\sqrt{2}}(\e_1-\e_2)$&3.22&3.26&3.22&0.0032\\ \hline
7&$\frac{1}{\sqrt{2}}(\e_1+\e_2)$&5.63&0.47&0.47& 0.0008\\ \hline
8&$\frac{1}{\sqrt{2}}(\e_1+\e_2)$&3.22&3.26&3.22&0.0032\\ \hline
9&$\frac{1}{\sqrt{2}}(\e_1-\e_3)$&0.55&5.67&0.55&0.00096\\ \hline
10&$\frac{1}{\sqrt{2}}(\e_1-\e_3)$&3.22&3.26&3.22& 0.0032\\ \hline
11&$\frac{1}{\sqrt{2}}(\e_1+\e_3)$&5.63&0.47&0.47& 0.0008\\ \hline 
12&$\frac{1}{\sqrt{2}}(\e_1+\e_3)$&3.22&3.26&3.22& 0.0032\\ \hline
\end{tabular}
\caption{The angles $\theta_1$, $\theta_2$ in degrees corresponding to Type I and Type II twins, and their minimum $\min\,(\theta_1,\theta_2)$, for different $k>2$, together with the values of $\delta_k=|\cof(\V_1-\V_k)|$. For $k=2$ there is a single angle $3.34^\circ$ (see above), while $\delta_2=0.0034$.}
\label{thetatable}
\end{table}
Note that the angles for $k=5,9$ and for $k=7,11$ are equal. This is because $|\U_1\e|$, $|\U_1^{-1}\e|$ are the same for $\e=\frac{1}{\sqrt{2}}(\e_1\pm\e_2)$ and  $\e=\frac{1}{\sqrt{2}}(\e_1\pm\e_3)$, while from  \eqref{aplus},\eqref{nplus} we have  $\ab_1\cdot\e_2=\ab_1\cdot\e_3$, $\n_1\cdot\e_2=\n_1\cdot\e_3$, $\n_1\cdot \U_1^{-2}\e_2=\n_1\cdot \U_1^{-2}\e_3$. Also the angles for $k=6,8,10,12$ are equal, the expression in \eqref{angleeven} taking the same value, as can be checked from \eqref{aplus},\eqref{nplus}.

That variants 5,7,9,11 are very nearly compatible with variant 1, as measured by the very small angles of rotation and low values of $\delta_k$, is a consequence of  $\frac{\ab_1}{|\ab_1|}$ being nearly parallel  to $\frac{\ab_k}{|\ab_k|}$ for $k=5,9$ and $\n_1$ being nearly parallel to $\n_k$ for $k=7,11$.

\begin{rem}
\label{measures}
See  \cite[Section 7]{DellaPorta19b} for a related study of different metrics for measuring how closely  the cofactor conditions are satisfied, in which a metric based on elastic stress is proposed for measuring incompatibility, and \cite{DellaPorta20} for  similar compatibility considerations related to non stress-free junctions.
\end{rem}

\section{$T_N$ configurations}
\label{TNsection}
  Following \cite{sverak95}, for $m\geq 1, n\geq 1$ and a compact subset $K$ of the space $\R^{m\times n}$ of real $m\times n$ matrices we define the {\it convex hull} $K^{\rm c}$, the {\it polyconvex hull},
the {\it quasiconvex hull} $K^{\rm qc}$,  and the {\it  rank-one convex hull} $K^{\rm rc}$ of $K$ by  
\begin{align*}K^{\rm c}&=\{\A\in \R^{m\times n}:\varphi(\A)\leq\max_K\varphi \text{ for all convex }\varphi:\R^{m\times n}\to\R\},\\K^{\rm pc}&=\{\A\in \R^{m\times n}:\varphi(\A)\leq \max_K\varphi\text{ for all polyconvex }\varphi:\R^{m\times n}\to\R\},\\
K^{\rm qc}&=\{\A\in \R^{m\times n}:\varphi(\A)\leq \max_K\varphi\text{ for all quasiconvex }\varphi:\R^{m\times n}\to\R\}, \\
K^{\rm rc}&=\{\A\in \R^{m\times n}:\varphi(\A)\leq \max_K\varphi\text{ for all rank-one convex }\varphi:\R^{m\times n}\to\R\},
\end{align*}
so that 
\be
\label{hullrelations}
K^{\rm rc}\subset K^{\rm qc}\subset K^{\rm pc}\subset K^{\rm c}.
\ee
See \cite{dacorogna} for the definitions of convex, polyconvex, quasiconvex and rank-one convex functions, together with equivalent forms of the above definitions. 

A {\it rank-one connection} in $K$ is a pair of matrices $\A,\B\in K$ with $\rank(\A-\B)=1$. If $\A,\B\in\R^{m\times n}$ then we denote by $[\A,\B]$ the closed line segment
$$[\A,\B]:=\{\lambda\A+(1-\lambda)\B:\lambda\in[0,1]\}.$$
A set $E\subset\R^{m\times n}$ is {\it rank-one convex} if $\A,\B\in E$ with $\rank(\A-\B)=1$ implies that $[\A,\B]\subset E$. If $K\subset\R^{m\times n}$ is compact then $K^{\rm rc}$ is rank-one convex, since for any rank-one convex function $\varphi$ and $\A,\B\in K^{\rm rc}$ with  $\rank(\A-\B)=1$, $\lambda\in[0,1]$, we have
$$\varphi(\lambda\A+(1-\lambda)\B)\leq \lambda \varphi(\A)+(1-\lambda)\varphi(\B)\leq \lambda \max_K\varphi+(1-\lambda)\max_K\varphi=\max_K\varphi.$$

If $\om\subset\R^n$ is bounded and open, a family of probability measures $(\nu_\x)_{\x\in\om}$ on $\R^{m\times n}$, depending measurably on $\x$, is {\it an $L^\infty$ gradient Young measure} if there is a sequence of continuous maps $\y^{(j)}:\om\to\R^m$ with weak derivatives $D\y^{(j)}$ bounded in $L^\infty(\om,\R^{m\times n})$ such that
$$f(D\y^{(j)})\weakstar \langle\nu_{\x},f\rangle:=\int_{\R^{m\times n}}f(\A)\,d\nu_\x(\A) \text{ in } L^\infty(\om)$$
for all continuous $f:\R^{m\times n}\to\R$, that is 
$$\int_\om f(D\y^{(j)})\,\theta(\x)\,d\x\to\int_\om\langle \nu_\x,f\rangle\,\theta(\x)\,d\x\text{ for all }\theta\in L^1(\om).$$
(For an intuitive probabilistic description of Young measures  see \cite{p21}.) Homogeneous $L^\infty$ gradient Young measures $\nu=\nu_\x$ (that is those that are independent of $\x$) are characterized (\cite{kinderlehrerpedregal91}, see also \cite{muller99}) by the two conditions (i)  $\nu$ has compact support in $\R^{m\times n}$ and (ii) the inequality 
\be
\label{gym}\langle\nu,\varphi\rangle\geq\varphi(\bar\nu)
\ee
holds for all quasiconvex $\varphi:\R^{m\times n}\to \R$, where $\bar \nu=\int_{\R^{m\times n}}\A\,d\nu(\A)$ denotes the centre of mass of $\nu$. Such a measure  is a {\it laminate} if \eqref{gym} holds for all rank-one convex $\varphi:\R^{m\times n}\to \R$. We will be particularly interested in {\it finite} homogeneous gradient Young measures and laminates, that is those that are finite convex combinations
$$\nu=\sum_{i=1}^N\lambda_i\delta_{\A_i},$$
of Dirac masses, where $\lambda_i\geq 0$, $\sum_{i=1}^N\lambda_i=1$, and $\A_i\in\R^{m\times n}$, which correspond to microstructures formed from gradients $\A_i$ with corresponding volume fractions $\lambda_i$.

We make frequent use of the following standard lemma.
\begin{lem}
\label{rankoneplusrankone}
Suppose that $\p\otimes\q$ and $\rb\otimes\s$ are  matrices in $\R^{m\times n}$ of rank $\leq 1$  such that
$$\p\otimes \q +\rb\otimes\s=\cb\otimes\db$$
for some $\cb\in\R^m, \db\in\R^n$.
Then either $\p$ is parallel to $\rb$ or $\q$ is parallel to $\s$.
\end{lem}
\begin{proof} For any $\vb$ with $\vb\cdot\db=0$ we have that $\p(\q\cdot\vb)+\rb(\s\cdot\vb)=\bzero$. Hence either $\p\parallel\rb$, or $\q\cdot\vb=\s\cdot\vb=0$ whenever $\vb\cdot\db=0$, in which case $\q\parallel\s$.
\end{proof}

Next we define $T_N$ configurations (see e.g., \cite[Def. 1]{szekelyhidi2005}):
\begin{defn}
\label{TNdefn}
An ordered set of $N\geq 1$ distinct matrices $\bigl\{\X_1,\dots,\X_N\bigr \} \subset\R^{m\times n}$ without rank-one connections forms a $T_N$ configuration if there exist $\Pb,\Cb_i\in\R^{m\times n}$ and real numbers $\kappa_i>1$ such that
\begin{align}
\X_1 &= \Pb + \kappa_1\Cb_1\nonumber\\
\X_2 &= \Pb + \Cb_1 + \kappa_2\Cb_2\nonumber\\
&\vdots\label{TNrel}\\
\X_N &= \Pb + \Cb_1 + \dots + \Cb_{N-1} + \kappa_N \Cb_N,\nonumber
\end{align}
and $\rank\Cb_i= 1$ and $\sum_{i=1}^N \Cb_i = \bzero.$
\end{defn}
\begin{figure}[h] 
  \centering
  \includegraphics[width=4.67in,height=3.13in,keepaspectratio]{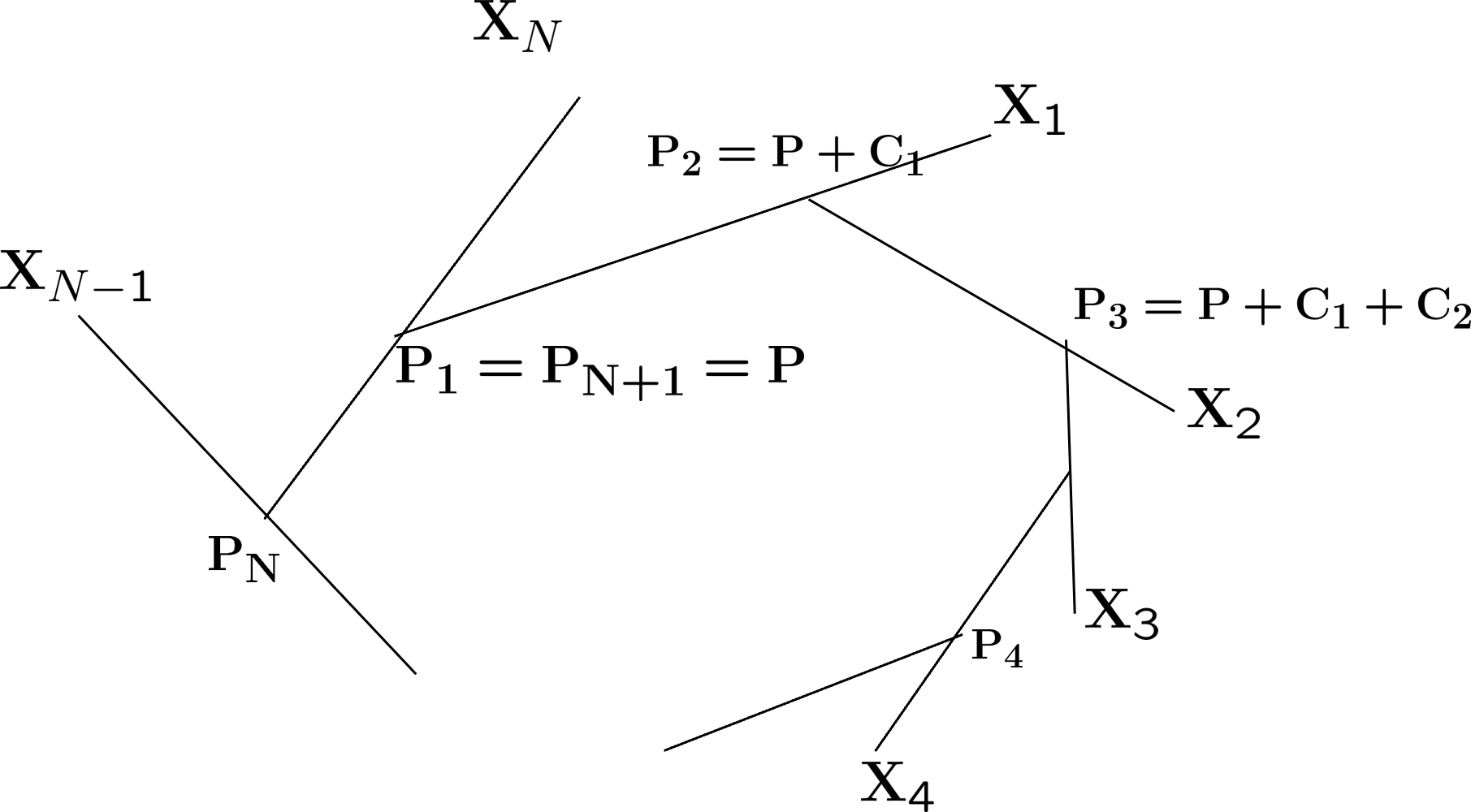}
  \caption{Schematic representation of $T_N$ configuration, with straight lines representing rank-one directions.}
  \label{fig:TNa}
\end{figure}
$T_N$ configurations are only possible if $N\geq 4$, since for $N=1$ we have $\Cb_1=\bzero$ and so $\rank\Cb_1\neq 0$, for $N=2$ \eqref{TNrel} implies that $\rank (\X_1-\X_2)=\rank((\kappa_1+\kappa_2-1)\Cb_1)= 1$, while if $N=3$ and $\Cb_1=\ab_1\otimes\n_1, \Cb_2=\ab_2\otimes\n_2$ the relation $\Cb_1+\Cb_2+\Cb_3=\bzero$ implies by Lemma \ref{rankoneplusrankone} that either $\ab_1\parallel\ab_2$ or $\n_1\parallel\n_2$, so that $\rank(\X_1-\X_2)\leq 1$.

The following result holds for $T_N$ configurations (see e.g., \cite[Lemma 2]{szekelyhidi2005} and \cite[Lemma 4.4]{mullersverak03}):
\begin{lem}
\label{TN}
Let $K=\{\X_1,\dots,\X_N\}\in \R^{m\times n}$ form a $T_N$ configuration. Then, setting
$$
\Pb_i = \Pb + \sum_{j<i}\Cb_j,
$$
the closed line segments $[\Pb_i,\X_i]$ are contained in $K^{\rm rc}$. Each point of $K^{\rm rc}$ is the centre of mass of a finite laminate supported on $K$; in particular such a finite laminate with centre of mass $\Pb$ is given by 
\be
\label{laminate}
\nu=\sum_{i=1}^N\theta_i\delta_{\X_i},
\ee
where $\beta_i=1-\frac{1}{\kappa_i}$ and
$$\theta_i:=\frac{(1-\beta_i)\prod_{j=1}^N\beta_j}{\prod_{j=1}^i\beta_j\left(1-\prod_{j=1}^N\beta_j\right)}.$$
\end{lem}
\begin{proof} 
We give the proof for the convenience of the reader, mainly following \cite{mullersverak03} for the case $N=4$. Let $\varphi:\R^{m\times n}\to\R$ be rank-one convex. For $1\leq i\leq N$ we have that, setting $\Pb_{N+1}:=\Pb_1=\Pb$, 
\be
\label{recursion}
\Pb_{i+1}=(1-\beta_i)\X_i+\beta_i\Pb_i,
\ee
and hence, since $\rank(\X_i-\Pb_i)=1$,
\be
\label{recursion1}
\varphi(\Pb_{i+1})\leq (1-\beta_i)\varphi(\X_i)+\beta_i\varphi(\Pb_i).
\ee
Multiplying \eqref{recursion} and \eqref{recursion1} by $\prod_{j=1}^N\beta_j/\prod_{j=1}^i\beta_j$ and summing from $i=1,\ldots,N$, noting that the terms involving $\Pb_2,\ldots,\Pb_{N}$ cancel, we obtain that
\begin{align}
\Pb&=\sum_{i=1}^N\theta_i\X_i\label{convcomb}\\
\varphi(\Pb)&\leq \sum_{i=1}^N\theta_i\varphi(\X_i).\label{r1ineq}
\end{align}
Since $\theta_i>0, \sum_{i=1}^N\theta_i=1$ it follows that $\varphi(\Pb)\leq \max_K\varphi$, and hence $\Pb\in K^{\rm rc}$. Also by \eqref{convcomb}, \eqref{r1ineq}, we have that $\nu$ in \eqref{laminate} satisfies $\Pb=\bar\nu$ and
$$\langle\nu,\varphi\rangle\geq\varphi(\Pb),$$
so that $\nu$ is a finite laminate. Since $K^{\rm rc}$ is rank-one convex, it follows from \eqref{recursion1} that $[\Pb_i,\X_i]\subset K^{\rm rc}$ for each $i$.

Finally, if $\A\in K^{\rm rc}$ then  a result of Kruzik \cite[Lemma 3]{kruzik} (for another proof see \cite{u5}) implies that $\A$ is a finite laminate supported on $K$.
\end{proof}
\begin{rem}
\label{lamination}
A generating sequence for $\nu$ can be given explicitly by a process of sequential lamination, as described for the case $N=4$ in \cite[Theorem 3.1]{Bhattetal94}.
\end{rem}
\begin{rem}
$K^{\rm rc}$ is in general larger than the union of the line segments $[\Pb_i,\X_i]$, as shown by the example due to Kirchheim \cite[Construction 4.34]{kirchheim03} (see also \cite{szekelyhidi2005}) of four matrices in $\R^{2\times 2}$ that form a $T_4$ configuration in different ways.
\end{rem}

We now follow the idea devised in \cite[Ex. 4.18, Cor. 4.19]{kirchheim03}.
\begin{defn}
An ordered set of $N\geq 1$ distinct matrices $\bigl\{\X_1,\dots,\X_N\bigr \} \subset\R^{m\times n}$ without rank-one connections forms a degenerate $T_N$ configuration if there exist $\Pb,\Cb_i\in\R^{m\times n}$ such that
\begin{align*}
\X_i &= \Pb + \Cb_i,\qquad\text{ for any $i=1,\dots,12$,}
\end{align*}
and $\rank \Cb_i= 1$, $\bzero\in\{\Cb_1,\ldots,\Cb_N\}^{\rm c}$.
\end{defn}
Degenerate $T_N$ configurations are also only possible for $N\geq 4$, since $\bzero=\sum_{i=1}^3\lambda_i\Cb_i$ for $\lambda_i\geq 0$ and $\sum_{i=1}^3\lambda_i=1$ implies by Lemma \ref{rankoneplusrankone} that two of the $\Cb_i$ are rank-one connected.
\begin{lem}
\label{deglem}
If $K=\{\X_1,\dots,\X_N\} \subset\R^{m\times n}$ forms a degenerate $T_N$ configuration then  $[\Pb,\X_i]\subset K^{\rm rc}$ for each $i$. If $\bzero=\sum_{i=1}^N\lambda_i\Cb_i$ for some $\lambda_i\geq 0$ with $\sum_{i=1}^N\lambda_i=1$, then 
$$\nu=\sum_{i=1}^N\lambda_i\delta_{\X_i}$$
is a laminate with centre of mass $\bar\nu=\Pb$.
\end{lem}
\begin{proof}
Since $\bzero\in\{\Cb_1,\ldots,\Cb_N\}^{\rm c}$ we have that $\bzero=\sum_{i=1}^N\lambda_i\Cb_i$ for some $\lambda_i\geq 0$ with $\sum_{i=1}^N\lambda_i=1$. We claim that $\Pb\in K^{\rm rc}$, from which  $[\Pb,\X_i]\subset K^{\rm rc}$ follows from the fact that $K^{\rm rc}$ is rank-one convex. By eliminating some $\X_i$ and reducing $N$ if necessary we may and will assume that each $\lambda_i>0$. 

For sufficiently small $\ep>0$ define $\X_i^\ep=\X_i+\ep\sum_{j=1}^i\lambda_j\Cb_j$. Then $\rank(\X_i^\ep-\X_j^\ep)>1$ for $i\neq j$,  and following \cite[Ex. 3]{szekelyhidi2005} we have that
\begin{align*}
\X_1^\ep&=\Pb+\left(1+\frac{1}{\ep\lambda_1}\right)\ep\lambda_1\Cb_1\\
\X_r^\ep&=\Pb+\sum_{j=1}^{r-1}\ep\lambda_j\Cb_j +\left(1+\frac{1}{\ep\lambda_r}\right)\ep\lambda_r\Cb_r,\;1<r\leq N,
\end{align*}
so that $\{\X_1^\ep,\ldots,\X_N^\ep\}$ forms a $T_{N}$ configuration, with $\X_N^\ep=\X_N$. Therefore, by Lemma \ref{TN},  $\Pb\in\{\X_1^\ep,\ldots,\X_N^\ep\}^{\rm rc}$.  Let $\varphi:\R^{m\times n}\to\R$ be rank-one convex (and thus continuous, see e.g. \cite[Lemma 5.6]{rindler2018}). Then 
$$\varphi(\Pb)\leq\max_j \varphi(\X_j^\ep),$$
so that letting $\ep\to 0$ we obtain $\varphi(\Pb)\leq \max_j\varphi(\X_j)$, and hence $\Pb\in K^{\rm rc}$ as claimed.

Now we apply the formula \eqref{laminate} to the $T_N$ configuration $\{\X_1^\ep,\ldots,\X_N^\ep\}$, noting that the corresponding $\beta_i^\ep=\frac{1}{1+\ep\lambda_i}$, and hence 
that if
$$ \lambda_i^\ep:=\frac{(1-\beta_i^\ep)\prod_{j=1}^N\beta_j^\ep}{\prod_{j=1}^i\beta_j^\ep\left(1-\prod_{j=1}^N\beta_j^\ep\right)}$$
then $\lim_{\ep\to 0+}\lambda_i^\ep=\lambda_i$.
Therefore if $\nu=\sum_{i=1}^N\lambda_i\delta_{\X_i}$ and $\varphi:\R^{m\times n}\to\R$ is rank-one convex we have that as $\ep\to 0+$
$$\varphi(\Pb)\leq \sum_{i=1}^N\lambda_i^\ep\varphi(\X_i^\ep)\rightarrow \sum_{i=1}^N\lambda_i\varphi(\X_i),$$
and hence $\nu$ is a laminate with $\bar\nu=\Pb$.
\end{proof}
\begin{rem}
The above proof, together with Remark \ref{lamination}, shows how a generating sequence of gradients $D\y^{(j)}$ for the laminate can be constructed, as well as justifying the terminology `degenerate $T_N$ configuration'. However there is a somewhat simpler proof pointed out to us by Bernd Kirchheim and suggested to him by Jan Kristensen, not involving $T_N$ configurations. Given a $C^1$ rank-one convex $\varphi$ we have that 
$$\varphi(\X_i)\geq \varphi(\bzero) +D\varphi(\bzero)\cdot\X_i,$$
so that if $\sum_{i=1}^N\lambda_i\X_i=\bzero$ and $\nu=\sum_{i=1}^N\lambda_i\delta_{\X_i}$ we obtain
\be
\label{r1cond}
\langle\nu,\varphi\rangle=\sum_{i=1}^N\lambda_i\varphi(\X_i)\geq \varphi(\bzero)=\varphi(\bar\nu).
\ee
Then \eqref{r1cond} follows for any rank-one convex (thus continuous) $\varphi$ by mollifying (which preserves rank-one convexity) and passing to the limit, so that $\nu$ is a laminate.
\end{rem}

\section{Analysis}
\label{analysis}
Let $m=n=3$, $\Z_i=\ab_i\otimes\n_i$ and $K=\{\Z_1,\ldots,\Z_N\}$ be defined as in \eqref{defK}, with $0<\alpha<1<\beta$. Throughout this section we assume that \eqref{nonpar} holds, that is
$2\alpha^2+\beta^2\neq 3,\, 2\alpha^{-2}+\beta^{-2}\neq 3$.
\begin{lem}
\label{Kpc1}
\begin{align}
\label{Kpc}
K^{\rm pc}&=K^{\rm c}\cap\{\ab\otimes\n:\ab,\n\in\R^3\},\\
 &\subset \{\ab\otimes\n:\ab,\n\in\R^3, \ab\cdot\n=\alpha\beta-1\}.\nonumber
\end{align}
\end{lem}
\begin{proof}
 We have that (see e.g.  \cite[Chapter 7]{dacorogna})
$$K^{\rm pc}=\{\A\in \R^{3\times 3}:T(\A)\in T(K)^{\rm c}\}$$
where $T(\B)=(\B,\cof\B,\det\B)$. Hence, since $\cof\B=\bzero$, $\det\B=0$ for $\B\in K$, $K^{\rm pc}=K^{\rm c}\cap\{\A\in \R^{3\times 3}:\cof\A=\bzero\}$,
giving the result since $\tr \A=\alpha\beta-1$ for all $\A\in K^{\rm c}$.
\end{proof}

\begin{prop}
\label{noTN}
There are no $T_N$ configurations in $K$. 
\end{prop}
\begin{proof}
Let $\{\hat\ab_1\otimes\hat\n_1,\ldots,\hat\ab_N\otimes\hat\n_N\}$ form a $T_N$ configuration in $K$ with corresponding point $\Pb$ as  in Definition \ref{TN}, so that the $\hat\ab_i\otimes\hat\n_i=\ab_{k(i)}\otimes\n_{k(i)}$ form an ordered subset of distinct points of  $K$. 

We first note that, by Lemma \ref{TN}, $\Pb\in K^{\rm rc}\subset K^{\rm pc}$, so that by Lemma \ref{Kpc1} $\Pb=\ab\otimes \n$ for some $\ab,\n\in\R^3$.
But also $\rank(\Pb-\hat\ab_1\otimes\hat\n_1)=\rank(\Pb-\hat\ab_N\otimes\hat\n_N)=1$, and so, by Lemma \ref{rankoneplusrankone}, $\ab\parallel\hat\ab_1$ or $\n\parallel\hat\n_1$, and $\ab\parallel\hat\ab_N$ or $\n\parallel\hat\n_N$. But, by Proposition \ref{atensorn}, $\hat\ab_1\not\parallel\hat\ab_N$ and $\hat\n_1\not\parallel\n_N$. Hence, either $\Pb=\lambda\hat\ab_1\otimes\hat\n_N$ or $\Pb=\lambda\hat\ab_N\otimes\hat\n_1$ for some constant $\lambda\neq 0$. Suppose that $\Pb=\lambda\hat\ab_1\otimes\hat\n_N$, the argument if $\Pb=\lambda\hat\ab_N\otimes\hat\n_1$ being similar. Then $$\Pb_2=\hat\ab_1\otimes((1-\frac{1}{\kappa_1})\lambda \hat\n_N+\frac{1}{\kappa_1}\hat\n_1)$$ But $\rank(\Pb_2-\hat\ab_2\otimes\hat\n_2)=1$, so that since $\hat\ab_1\not\parallel\hat\ab_2$ the vectors $\hat\n_N,\hat\n_1,\hat\n_2$ are collinear, and thus $[\hat\n_N,\hat\n_1,\hat\n_2]=0$. Hence by Proposition \ref{collinear}(ii) we have that $\alpha^2+2\beta^2=3$. Similarly, since $$(\kappa_N-1)\Pb_N=(\kappa_N\lambda\hat\ab_1-\hat\ab_N)\otimes \hat\n_{N-1}$$ and $\rank(\Pb_N-\hat\ab_{N-1}\otimes \hat\n_{N-1})=1$ we have that $[\hat\ab_{N-1},\hat\ab_N,\hat\ab_1]=0$. Hence by  Proposition \ref{collinear}(i) we have that $\alpha^{-2}+2\beta^{-2}=3$, which together with $\alpha^2+2\beta^2=3$ implies $\alpha=1$, a contradiction.
\end{proof}
\begin{lem}
\label{Pzero}
If there exists a degenerate $T_N$ configuration in $K$ then the corresponding $\Pb=\bzero$. If $\alpha\beta\neq 1$ then there is no degenerate $T_N$ configuration in $K$.
\end{lem}
\begin{proof}
Without loss of generality we may suppose that  $\{\ab_1\otimes \n_1,\ldots, \ab_N\otimes\n_N\}$ is a degenerate $T_N$ configuration in $K$. By Lemma \ref{deglem}, $\Pb\in K^{\rm rc}\subset K^{\rm pc}$. Hence by Lemma \ref{Kpc1} we have that $\Pb=\ab\otimes\n$ for some $\ab,\n$ with $\ab\cdot\n=\alpha\beta-1$. By Lemma \ref{rankoneplusrankone}, $\ab\parallel\ab_i$ or $\n\parallel\n_i$ for $1\leq i\leq N$. Since $N\geq 4$ this implies that if $\Pb\neq\bzero$ then two of the $\ab_i$ or two of the $\n_i$ are parallel, which is impossible by Proposition \ref{atensorn}. Hence $\Pb=\bzero$, which is impossible if $\alpha\beta\neq 1$.
\end{proof}
\begin{prop}
\label{LemmaRc}
Let $\alpha\beta=1$. 

\noindent (i) There exists a degenerate $T_{12}$ configuration in $K$, and for any $t\in[0,1]$ the set $tK\subset K^{rc}$.

\noindent (ii)  Assume $1<\beta<\sqrt{2}$. If $\kappa_i\geq 0, i=1,2,3,4,$ with $\sum_{i=1}^4\kappa_i=1$ and $\lambda_j=\frac{1}{12}\sum_{i=1}^4\kappa_ib^{(i)}_j$ (where the $\bb^{(i)}$ are defined as in Lemma \ref{nullE}) then $\nu=\sum_{j=1}^{12}\lambda_j\delta_{\ab_j\otimes\n_j}$ is a laminate supported in $K$ with centre of mass $\bzero$. For any such laminate 
\be
\label{thirds}
\sum_{j=1}^4\lambda_j=\sum_{j=5}^8\lambda_j=\sum_{j=9}^{12}\lambda_j=\frac{1}{3}.
\ee
\end{prop}
\begin{proof}
 By Proposition \ref{atensorn} there are no rank-one connections in $K$, while by  Lemma \ref{sumrank} we have that $\bzero\in K^{\rm c}$. Hence there is a degenerate $T_{12}$ configuration in $K$ with $\Pb=\bzero$, and $\bzero\in K^{\rm rc}$ by  Lemma \ref{deglem}. Since $K^{\rm rc}$ is rank-one convex it follows that $tK\subset K^{rc}$.

	To prove (ii) note that $\beta\leq \sqrt{2}$ implies $\tau\leq 1$ and hence $b^{(i)}_j\geq 0$ for all $i,j$  and $\lambda_j\geq 0$ for all $j$.  Since each $\bb^{(i)}\in N(\E)$ we have that $\sum_{j=1}^{12}\lambda_j\ab_j\otimes\n_j=\bzero$, so that by Lemma \ref{deglem} $\nu=\sum_{j=1}^{12}\lambda_j\delta_{\ab_j\otimes\n_j}$ is a laminate supported in $K$ with centre of mass $\bzero$. Also, since $\sum_{j=1}^{4}b^{(i)}_j=\sum_{j=5}^{8}b^{(i)}_j=\sum_{j=9}^{12}b^{(i)}_j=4$ for each $i$ and the $\bb^{(i)}$ span $N(\E)$ we have that\eqref{thirds} holds for any such laminate.
\end{proof}
\begin{rem}
\label{volumefracs}
In addition to \eqref{thirds} there are other relations between the volume fractions $\blambda=(\lambda_1,\ldots,\lambda_{12})$ of laminates $\nu=\sum_{j=1}^{12}\lambda_j\delta_{\ab_j\otimes\n_j}$ supported in $K$ with centre of mass $\bzero$. If $\z=(z_1,\ldots,z_{12})$ belongs to the 8-dimensional orthogonal complement of $N(\E)={\rm span}\,\{\bb^{(1)},\bb^{(2)},\bb^{(3)},\bb^{(4)}\}$ in $\R^{12}$ then $\z\cdot\blambda=0$. For example we can take \begin{align*}
\z^{(1)}&=(1,0,0,1,-1,0,-1,0,0,0,0,0),\\ \z^{(2)}&=(0,1,1,0,0,-1,0,-1,0,0,0,0),
\end{align*}
giving 
\be
\label{lrelations}
\lambda_1+\lambda_4=\lambda_5+\lambda_7,\;\lambda_2+\lambda_3=\lambda_6+\lambda_8,
\ee
and similarly
\begin{align}
\label{lrelations1}\lambda_1+\lambda_3&=\lambda_9+\lambda_{11},\;\lambda_2+\lambda_4=\lambda_{10}+\lambda_{12},\\
\lambda_5+\lambda_8&=\lambda_9+\lambda_{12},\;\lambda_6+\lambda_7=\lambda_{10}+\lambda_{11},
\label{lrelations2}
\end{align}
from which also \eqref{thirds} follows.
\end{rem}
In particular, taking $\kappa_1=\kappa_2=\kappa_3=0, \kappa_4=1$ in (ii), we see  that $\sum_{j=1}^{12}\frac{1}{12}\delta_{\ab_j\otimes\n_j}$ is a laminate supported in $K$ with centre of mass $\bzero$ and equal volume fractions $\lambda_j=\frac{1}{12}$. But other choices of the $\kappa_j$ give rise to such  laminates with unequal volume fractions satisfying \eqref{thirds}-\eqref{lrelations2}.

{\it We are not able to determine} $K^{\rm qc}$. A conjecture is that $K^{\rm qc}=[0,1]K$ when $\alpha\beta=1$. In related problems a standard proof technique is to show that $K^{\rm rc}=K^{\rm pc}$ so that $K^{\rm qc}=K^{\rm rc}$ by \eqref{hullrelations}. However, if $\alpha\beta=1$, since by Lemma \ref{subsumrank} (i) $\bzero$ is an interior point of $K^{\rm c}$ in the subspace $\{\A\in\R^{3\times 3}:\tr\A=0\}$, it follows from Lemma \ref{Kpc1} that for $|\ep|$ sufficiently small $K^{\rm pc}\supset \{\ab\otimes \n:|\ab|+|\n|<\ep, \ab\cdot\n=0\}$, so that $K^{\rm pc}$ contains points not in $[0,1]K$ and this method does not work. 

In the case $\alpha\beta\neq 1$ a conjecture (supported to some extent by Lemma \ref{Pzero} and Proposition \ref{noTN}) is that $K^{\rm qc}=K$. Here also $K^{\rm pc}$ is strictly bigger than $K$, at least when $0<\bar\alpha<1<\bar\beta$ with $\bar\alpha\bar\beta=1$ and $|\alpha-\bar\alpha|+|\beta-\bar\beta|$ is sufficiently small. To see this write $K=K(\alpha,\beta)$ and note that $K(\alpha,\beta)$ depends continuously on $(\alpha,\beta)$ in the Hausdorff metric of $\R^{3\times 3}$ in a neighbourhood of $(\bar\alpha,\bar\beta)$ and so also does $K(\alpha,\beta)-\frac{1}{3}(\alpha\beta-1)\1\subset\{\A\in\R^{3\times 3}:\tr \A=0\}$. Therefore $K(\alpha,\beta)^{\rm c}-\frac{1}{3}(\alpha\beta-1)\1=(K(\alpha,\beta)-\frac{1}{3}(\alpha\beta-1)\1)^{\rm c}\to K(\bar\alpha,\bar\beta)^{\rm c}$ in the Hausdorff metric of $\{\A\in\R^{3\times 3}:\tr \A=0\}$ as $(\alpha,\beta)\to(\bar\alpha,\bar\beta)$. Hence by  Lemma \ref{subsumrank} there exists $\ep>0$ such that for all sufficiently small  $|\alpha-\bar\alpha|+|\beta-\bar\beta|$ we have $B(\bzero,\ep)\cap \{\A\in\R^{3\times 3}:\tr \A=0\}\subset K(\alpha,\beta)^{\rm c}-\frac{1}{3}(\alpha\beta-1)\1$. Hence $(\alpha\beta-1)\n\otimes\n\in K(\alpha,\beta)^{\rm pc}$ for all unit vectors $\n$, and  thus $K(\alpha,\beta)^{\rm pc}$ contains points not in $K$.

Next we explore the existence of degenerate $T_N$ configurations in $K$  with $1\leq N<12$. 
Since $\bzero$ is in the convex hull of $K$, which is a subset of the 8-dimensional subspace $\{\A\in \R^{3\times 3}:\tr \A=0\}$, by Carath\'eodory's theorem $\bzero =\sum_{i=1}^9 \lambda_i\ab_{m(i)}\otimes \n_{m(i)}$, where $\lambda_i\geq 0$, $\sum_{i=1}^9\lambda_i=1$ and the $m(i)\in\{1,2,\ldots,12\}$ are distinct. Hence there exist $T_9$ configurations in $K$. However more is true.
\begin{prop}
\label{otherTN}Let $\alpha\beta=1$, $1<\beta<\sqrt 2$.

\noindent(i) there exist degenerate $T_8$ configurations in $K$,

\noindent (ii) if $N<8$ then $\sum_{j=1}^N\lambda_j\ab_{m(j)}\otimes\n_{m(j)}\neq \bzero$ for any convex combination of elements of $K$ with $\lambda_j>0$ for $1\leq j\leq N$; in particular there is no degenerate $T_N$ configuration in $K$ for $N<8$.

\noindent(iii) If for some $N<12$, $\nu=\sum_{j=1}^N\lambda_j\delta_{\ab_{m(j)}\otimes \n_{m(j)}}$ is a laminate with centre of mass $\Pb=\bzero$ and $\lambda_j>0$ for $1\leq j\leq N$ then the $\lambda_i$ are not all equal.
\end{prop}
\begin{proof}(i) For each $i=1,2,3$,  $\frac{1}{12}\sum_{j=1}^{12}b^{(i)}_j\ab_j\otimes\n_j=\bzero$, and since $b^{(i)}_j\geq 0$ for each $j$, $\frac{1}{12}\sum_{j=1}^{12} b^{(i)}_j =1$, and  four of the $b^{(i)}_j$ are zero these give degenerate $T_8$ configurations in $K$ with $\Pb=\bzero$. 

\noindent (ii) We have to prove that there is no nonzero linear combination $\x=\sum_{i=1}^4\alpha_i\bb^{(i)}$ of the basis vectors $\bb^{(i)}$ with each $x_j\geq 0$ and five or more of the $x_j$ zero. The $x_j$ are given explicitly by
\begin{align*}
x_1&=(2,1-\tau,2,1)\cdot\balpha,&
x_2&=(0,1-\tau,0,1)\cdot\balpha,\\
x_3&=(0,1+\tau,2,1)\cdot\balpha,&
x_4&=(2,1+\tau,0,1)\cdot\balpha,\\
x_5&=(2,2,1-\tau,1)\cdot\balpha,&
x_6&=(0,0,1-\tau,1)\cdot\balpha,\\
x_7&=(2,0,1+\tau,1)\cdot\balpha,&
x_8&=(0,2,1+\tau,1)\cdot\balpha,\\
x_9&=(1-\tau,2,2,1)\cdot\balpha,&
x_{10}&=(1-\tau,0,0,1)\cdot\balpha,\\
x_{11}&=(1+\tau,0,2,1)\cdot\balpha,&
x_{12}&=(1+\tau,2,0,1)\cdot\balpha,
\end{align*}
where $\balpha=(\alpha_1,\alpha_2,\alpha_3,\alpha_4)$. Without loss of generality it is sufficient to consider the following three cases: (a) $x_9=x_{10}=x_{11}=x_{12}=0$, (b) exactly three of $x_9,x_{10},x_{11},x_{12}$ and two other $x_i$ are zero, (c) exactly two of $x_9,x_{10},x_{11},x_{12}$, exactly two of $x_5,x_6,x_7,x_8$ and one of $x_1,x_{2},x_{3},x_{4}$ are zero. The easiest case is (a), since then \eqref{det} implies that $\balpha=\bzero$. The next easiest case is (b), when we have that $\balpha=\alpha_1(1,-\tau,-\tau,\tau-1)$ or $\alpha_1(1,\tau,\tau,-1-3\tau)$ or $\alpha_1(1,\tau,-\tau,\tau-1)$ or $\alpha_1(1,-\tau,\tau,\tau-1)$, and it is easily checked that then since $\tau>0, \tau\neq 1$ none of $x_1,\ldots, x_{8}$ is zero for $\alpha_1\neq 0$. 

For the case (c) there are six choices of pairs from each of $x_{9},x_{10},x_{11},x_{12}$ and $x_5,x_6,x_7,x_8$. A lengthy calculation shows that the only compatible options with $\balpha\neq\bzero$ are:
\begin{align*}
x_9=x_{10}=x_5=x_6=0,\;\;  &\balpha=\alpha_1(1,-1,1,\tau-1),\\
x_9=x_{10}=x_7=x_8=0,\;\; &\balpha=\alpha_1(1,1,-1,\tau-1),\\
x_9=x_{12}=x_5=x_8=0,\;\;&\balpha=\alpha_2(0,1,0,-2),\\
x_9=x_{11}=x_5=x_7=0,\;\;&\balpha=\alpha_1(1,\tau,1,-3-\tau),\\
x_9=x_{11}=x_6=x_8=0.\;\;&\balpha=\alpha_1(1,\tau,-1,1-\tau),\\
x_{10}=x_{12}=x_5=x_7=0,\;\;&\balpha=\alpha_1(1,-\tau,-1,\tau-1),\\
x_{10}=x_{11}=x_6=x_7=0,\;\;&\balpha=\alpha_2(0,1,0,0),\\
x_{12}=x_{11}=x_5=x_6=0,\;\;&\balpha=\alpha_1(1,-1,-1,1-\tau),\\
x_{12}=x_{11}=x_7=x_8=0,\;\;&\balpha=\alpha_1(1,1,1,-3-\tau).
\end{align*}
Checking whether these values of $\balpha$ can give rise to one of $x_1,x_{2},x_{3},x_{4}$ being zero we find that the only possibilities are $\balpha=\alpha_1(1,-1,1,\tau-1)$, $\balpha=\alpha_1(1,-1,-1,1-\tau)$, $\balpha=\alpha_1(1,1,-1,\tau-1)$ and $\balpha=\alpha_1(1,1,1,-3-\tau)$, which give respectively
\begin{align*}\x&=2\alpha_1(1+\tau,\tau-1,0,0,0,0,1+\tau,\tau-1,0,0,1+\tau,\tau-1)^T\\
\x&=2\alpha_1(0,0,-1-\tau,1-\tau,0,0,1-\tau,-1-\tau,-1-\tau,1-\tau,0,0)^T\\
\x&=2\alpha_1(0,0,\tau-1,1+\tau,1+\tau,\tau-1,0,0,0,0, \tau-1,1+\tau)^T\\
\x&=2\alpha_1(1-\tau,-1-\tau,0,0,1-\tau,-1-\tau,0,0,1-\tau,-1-\tau,0,0)^T.
\end{align*}
Since $0<\tau<1$ it is impossible to choose $\alpha_1\neq 0$ in any of these cases so that $x_j\geq 0$ for all $j$.

That there is no degenerate $T_N$ configuration in $K$ for $N<8$ follows by Lemma \ref{Pzero}.

\noindent (iii) This follows from Lemma \ref{subsumrank} (ii).
\end{proof}

 Now let $\om\subset\R^3$ be a bounded Lipschitz domain  with boundary $\partial\om$, and consider the Sobolev space
$W^{1,\infty}(\om;\R^3)$ of bounded measurable maps $\z:\om\to\R^3$ with essentially bounded weak derivative $D\z(\x)$ and norm
$$\|\z\|_{1,\infty}:= \esup_{\x\in\om}(|\z(\x)|+|D\z(\x)|).$$
Since $\om$ is bounded and Lipschitz every $\z\in W^{1,\infty}(\om;\R^3)$ has a representative (which we choose) that is continuous on the closure $\overline\om$ of $\om$ (see, for example, \cite[Theorem 4.12]{adamsfournier03}), so that in particular $\z(\partial\om)$ is well defined.

The following result is  useful: 
\begin{prop}
\label{2dim}
Let $\z \in W^{1,\infty}(\Omega;\R^3)$ be such that $\rank \,D\z\leq 1$ a.e. in $\Omega$ and either $\Div \z \geq 0$ a.e. in $\Omega$ or $\Div\z \leq 0$  a.e. in $\Omega$. Then  $\z (\om)\subset \z(\partial\om)$.  
\end{prop}
\begin{proof}
The proof follows closely the argument of \cite[Prop. 2.6]{kirchheim03}, which uses a key idea from \cite{chlebik}, but we give the details for completeness. 
We assume first that $\Div \z\geq 0$ and for $\delta>0$ define $\z_\delta(\x): = \delta \vc \x + \vc z(\x)$. Then
$$
D \z_\delta = \delta\1 + D\vc z,\quad
\det D \z_\delta = \delta^{2} (\delta + \Div \vc z)\geq \delta^3>0.$$ 
As a consequence, $\z_\delta$ is a map of bounded distortion and by \cite[Thm. 6.4 in Chap. II]{reshetnyak89} $\z_\delta$ is also an open map. As a consequence $\partial \z_\delta (\om) \subset \z_\delta (\partial \om)$. Suppose there exists some $\p \in \vc z(\om)\setminus \vc z(\partial\om)$, and set $\dist (\p,\z(\partial\om))=2\ep>0$. As the convergence of $\z_\delta$ to $\vc z$ is uniform as $\delta\to 0,$ $B(\p,\eps)\cap \partial \z_\delta(\om) \subset B(\p,\ep)\cap\z_\delta(\partial\om) =\varnothing$ for all sufficiently small $\delta$. Since $\p=\z(\x)$ for some $\x\in\om$  the uniform convergence implies that $B(\p,\eps)\cap \z_\delta (\om) \neq \varnothing$ for sufficiently small $\delta $ and thus that $B(\p,\eps)\subset \z_\delta (\om)$. In particular, denoting by $|E|$ the volume of an open set $E\subset\R^3$, we have  by the area formula (see, for example, \cite[Theorem 5.6]{fonsecagangbo95a}) that
$$
|B(\p,\eps)| \leq |\z_\delta (\om)| \leq \int_\om \det D\z_\delta(\x)\,d\x\leq c\delta^2 |\Omega| 
$$ 
for some $c>0$ depending only on $\om$ and $\|\vc z\|_{1,\infty}$. Then letting $\delta\to 0$  we get a contradiction. 

If $\Div\z\leq 0$ we take $\delta<0$, so that $\det D \z_\delta\leq -\delta^3<0$, and follow the same argument. 
\end{proof}
As a consequence we can prove the following result:
\begin{thm}
\label{Corrigid}
Let $\vc z \in W^{1,\infty}(\Omega;\R^3)$ be such that $\rank D\vc z \leq 1$  and $\Div \vc z = \xi$ a.e. in $\Omega$  for some $\xi\in\R.$ If $\vc z(\x)=\A\x+\bb$  for all $\x\in\partial\Omega$, where $\A\in \R^{3\times 3}, \bb\in\R^3$, then $\A=\ab\otimes\n$ for $\ab,\n\in\R^3$ with $\ab\cdot\n=\xi$, and $\vc z(\x)=\A\x+\bb$ for all $\x\in\Omega$. 
\end{thm}
\begin{proof}
The proof  follows closely that of \cite[Cor. 2.7]{kirchheim03}.   Since the minors are null Lagrangians we have 
$$
\cof\A = \frac{1}{|\Omega|}\mint\cof D\vc z(\vc x)\,\mathrm d\vc x = \bzero,
$$
from which we deduce that $\A = \vc a\otimes\vc n$ for some $\vc a,\vc n\in\R^3.$ Therefore
\begin{align}
\z(\partial\om)&=\{\A\x+\bb:\x\in\partial\om\}\nonumber\\
&=\{\ab(\x\cdot\n)+\bb:\x\in\partial\om\}.
\label{boundaryaffine}
\end{align}
 In the same way we deduce that $\tr \A=\ab\cdot\n=\xi$. Then, by Proposition \ref{2dim} and \eqref{boundaryaffine} we have that  $\vc z(\x) = f(\vc x)\vc a+\bb$, for some $f\in W^{1,\infty}(\Omega;\R)$. Therefore $\hat{\vc z}(\x):=\vc z(\x)- \A\vc x - \vc b$ satisfies $D\hat\z(\x)=\ab\otimes(\nabla f(\x)-\n)$, so that $\rank D\hat{\vc z}\leq 1$ a.e., $\Div\hat\z=0$ a.e. and $\hat{\vc z} = \vc 0$ on $\partial\Omega$. Therefore another application of Proposition \ref{2dim} implies that $\hat{\vc z} = \vc 0$ in $\Omega,$ giving the result.
\end{proof}
\begin{cor}
\label{rigidv} If $\vc y\in W^{1,\infty}(\om;\R^3)$ with $D\vc y(\x)\in H^{\rm pc}$  for a.e. $\x\in\om$ and $\vc y(\x)=\vc A\x$ for $\x\in\partial\om$ for some $\vc A\in \R^{3\times 3}$, then $\vc A=\1+\vc a\otimes \vc n$ with $\vc a\cdot\vc n=\alpha\beta-1$ and $\vc y(\x)=\vc A\x$ for a.e. $\x\in\om$. If further $D\vc y(\x)\in H$ for a.e. $\x\in\om$ then $\vc A\in H$.
\end{cor}
\begin{proof}  Let $\vc z(\x)=\vc y(\x)-\x$. Then $D\vc z(\x)\in K^{\rm pc}$ for a.e. $\x\in\om$, so that, by Lemma \ref{Kpc1},  $\Div \vc z(\x)=\alpha\beta-1$ for a.e. $\x\in\om$. Let $\varphi:\R^{3\times 3}\to\R$ be polyconvex. Then $\varphi$ is quasiconvex and so
$$\varphi(\vc A-\1)\leq \av_\om\varphi(D\vc z(\x))\,d\x\leq\max _{K^{\rm pc}}\varphi=\max_K\varphi.$$
Therefore $\vc A-\1\in K^{\rm pc}$, so that by Lemma \ref{Kpc1} we have that $\vc A=\1+\vc a\otimes \vc n$ with $\vc a\cdot\vc n=\alpha\beta-1$. Theorem \ref{Corrigid}  then implies that  $\vc z(\x)=(\vc A-\1)\x$ for a.e. $\x\in\om$, so that $\vc y(\x)=\vc A\x$ for a.e. $\x\in\om$. Hence also if $D\vc y(\x)\in H$ for a.e. $\x\in\om$ then $\vc A\in H$.
\end{proof}
Despite this rigidity result, one can solve the differential inclusion $D\vc u\in U$ with suitable  boundary conditions for an arbitrarily small neighbourhood $U$ of $K$ as a consequence of the following result due to M\"uller and \v{S}ver\'{a}k in \cite{MullerSverak99} (see also \cite[Thm 1.1]{MullerSverak}):
\begin{thm}(\cite[Thm. 1.1]{MullerSverak99})
\label{MS THM}
Let $U\subset \R^{3\times 3}$ be bounded and open. Let $\vc v\colon\Omega\subset\R^3\to\R^3$ be a piecewise linear Lipschitz, or $C^1$, map that satisfies $D\vc v(\x)\in U^{rc}$ for a.e. $\x\in\Omega$. Then there exists a Lipschitz map $\vc u:\Omega\to\R^3$ such that $D\vc u\in U$ a.e. in $\Omega$ and $\vc u=\vc v$ on $\partial\Omega$.
\end{thm}
Here $U^{\rm rc}$ is defined to be the union of the hulls $D^{\rm rc}$ for $D\subset U$ compact. For example if $\alpha\beta=1$ we can take $U$ to be the $\ep-$neighbourhood $N_\ep(K):=\{\B\in\R^{3\times 3}:|\B-\A|<\ep \text{ for some }\A\in K\}$ of $K$ and $\vc v$ a piecewise linear or $C^1$ map satisfying $|D\vc v(\x)|<\ep$ for a.e. $\x\in\om$, since by Proposition \ref{LemmaRc}(i) we have that  $\bzero\in K^{\rm rc}$ and thus $\A\in (K+\A)^{\rm rc}\subset U^{\rm rc}$ if $|\A|<\ep$.

\section{Discussion}
\label{Discussion}
In this section we discuss the implications of our analysis for the observed microstructures in \Tn. We consider a single grain occupying the bounded Lipschitz domain $\om\subset\R^3$ in the reference configuration, and take the point of view, supported by the experimental evidence in Section \ref{expresults},  that because $\lambda_2(\U_i)=1$ the austenite-martensite transition takes place via the nucleation of  exact austenite-martensite interfaces. These are described by the rank-one connections $\1+\ab_i\otimes\n_i$ between $\1$ and $\cup_{i=1}^6SO(3)\U_i$ that  comprise the set $H=\1+K$ defined in \eqref{defH}.
\subsection{Exact gradients and zero-energy microstructures}
The experimental observations suggest that the deformation $\y:\om\to\R^3$ of the grain is such that $D\y(\x)$ is very close to $H$ except perhaps for $\x$ belonging to a subset of $\om$ of very small volume. So we first consider the `exact' case of deformations $\y:\om\to\R^3$ satisfying
\begin{equation}
\label{exactgradient}
D\y(\x)\in H \text{ for a.e. }\x\in\om,
\end{equation}
and of corresponding zero-energy microstructures supported on $H$, that is of gradient Young measures $(\mu_\x)_{\x\in\om}$ satisfying
\begin{equation}
\label{GYM}
\supp \mu_\x\subset H\text{  for a.e. }\x\in\om,
\end{equation}
with underlying macroscopic deformation $\y:\om\to\R^3$ satisfying $D\y(\x)\in H^{\rm qc}$ for a.e. $\x\in\om$.  We note that, assuming the free-energy density $\psi(\cdot,\theta)$ to be $C^1$, we have that $D_\A\psi(\A,\theta)=\bzero$ for $\A\in H$, so that deformations $\y$ satisfying \eqref{exactgradient} are stress-free. Similarly, for gradient Young measures satisfying \eqref{GYM} the microscopic stress is zero, so that also the  macroscopic stress $\langle\mu_\x, D_\A\psi(\cdot,\theta)\rangle:=\int_{\supp \mu_{\x}}D\psi(\A,\theta)\,d\mu_\x(\A)=\bzero$ for a.e. $\x\in\om$.

In terms of the displacement $\z(\x):=\y(\x)-\x$, \eqref{exactgradient} is equivalent to 
\begin{equation}
\label{exactgradientz}
D\z(\x)\in K \text{ for a.e. }\x\in\om,
\end{equation}
while setting $\nu_x(E)=\mu_x(E+\1)$ for $E\subset\R^{3\times 3}$, \eqref{GYM} is equivalent to considering gradient Young measures $(\nu_\x)_{\x\in\om}$  satisfying
\begin{equation}
\label{GYMz}
\supp \nu_\x\subset K\text{  for a.e. }\x\in\om,
\end{equation}
with underlying macroscopic deformation $\z:\om\to\R^3$ satisfying $D\z(\x)=\bar\nu_\x\in K^{\rm qc}$ for a.e. $\x\in\om$.

  To complete the model we need to add a boundary condition. We first suppose that the grain boundary is constrained by the neighbouring grains so that it deforms only by the dilatation having determinant that of the matrices $\1+\ab_i\otimes\n_i$, that is (assuming without loss of generality that $\bzero\in\om$ and $\y(\bzero)=\bzero$)
\begin{equation}
\label{bc}
\y(\x)=(\alpha\beta)^\frac{1}{3}\x \text{ for all } \x\in\partial\om,
\end{equation}
so that $\z(\x)=((\alpha\beta)^\frac{1}{3}-1)\x$ for $\x\in\partial\om$, and thus $\left.\z\right|_{\partial\om}=\bzero$ if $\alpha\beta=1$. As explained in Section \ref{expresults} \eqref{bc} is consistent with the observations of the grain boundary. Furthermore \eqref{bc} is a consistent boundary condition for the polycrystal as a whole, since a uniform dilatation of the whole polycrystal gives the same dilatation for each grain. A more general analysis would have to consider compatibility at the grain boundary (for a recent study see \cite{u21}).  

\subsection{The case $\alpha\beta=1$.}
\label{6.2}
Suppose that $\alpha\beta=1$. Then since $\bzero\not\in K$, by Corollary \ref{rigidv} with $\vc A=\1$ there is no $\y\in W^{1,\infty}(\om;\R^3)$ satisfying \eqref{exactgradient} and \eqref{bc}, so that there is  no deformation using just the twelve deformation gradients observed experimentally which does not deform the grain boundary. It is possible that more is true, {\it that any $\y\in W^{1,\infty}(\om;\R^3)$ satisfying \eqref{exactgradient} is such that $D\y(\x)$ is constant,} but this seems difficult to decide.

On the other hand, by Proposition \ref{LemmaRc} (ii), if $\kappa_i\in L^\infty(\om), \, i=1,\ldots,4$ with $\kappa_i\geq 0,\,\sum_{i=1}^4\kappa_i=1$, and the $\bb^{(i)}$ are as in Lemma \ref{nullE} then 
\begin{equation}
\label{nu}
\nu_\x=\frac{1}{12}\sum_{j=1}^{12}\sum_{i=1}^4 \kappa_i(\x)b^{(i)}_j\delta_{\ab_j\otimes\n_j}
\end{equation}
is a solution to \eqref{GYMz} with underlying macroscopic deformation $\z=\bzero$. This follows because $\nu_\x$ is a laminate with $\bar\nu_\x=\bzero$ for each $\x$ and the characterization of $W^{1,\infty}$ gradient Young measures due to Kinderlehrer \& Pedregal \cite[Theorem 6.1]{kinderlehrerpedregal91}. 

A possible choice of the $\kappa_i$ is $\kappa_1=\kappa_2=\kappa_3=0,\,\kappa_4=1$, when $\nu:=\nu_\x=\frac{1}{12}\sum_{j=1}^{12}\delta_{\ab_j\otimes\n_j}$ is a homogeneous gradient Young measure with equal volume fractions $\lambda_j=\frac{1}{12}$. Another possible choice is $\kappa_i=1$ for some $i\in\{1,2,3\}$, $\kappa_j=0$ for $j\neq i$, leading to a homogeneous gradient Young measure $\nu=\frac{1}{12}\sum_{j=1}^{12}b^{(i)}_j\delta_{\ab_j\otimes\n_j}$ supported on just 8 of the $\ab_j\otimes\n_j$, and with corresponding volume fractions that are not all equal. Further, by Proposition \ref{otherTN}(iii), any homogeneous gradient Young measure supported on less than 12 of the $\ab_j\otimes\n_j$ and with underlying macroscopic deformation $\z=\bzero$ has  volume fractions that are not all equal. More generally, \eqref{nu} shows that there are gradient Young measures supported on $K$, and with underlying macroscopic deformation $\z=\bzero$, having volume fractions that vary with $\x$ across the grain. These microstructures are infinitely fine, with at least 8 values of the gradient used at each point $\x$.  In fact in Figure 3 one can observe regions dominated by just 8 of the variants.

Note also that if 
$$(\nu_{\x})_{\x\in\om}=\sum_{j=1}^{12}\lambda_j(\x)\delta_{\V_j}$$ is any gradient Young measure supported a.e.  on $K$, and with underlying macroscopic deformation $\y$ satisfying $\y(\x)=\x$ on $\partial\om$, then 
$$\nu:=\av_\om\nu_{\x}\,d\x=\sum_{j=1}^{12}\left(\av_{\om}\lambda_j(\x)\,d\x\right)\delta_{\ab_j\otimes\n_j}$$
is a homogeneous gradient Young measure satisfying $\bar\nu=\bzero$. Hence the volume fractions  $\lambda_j:=\smallav_{\om}\lambda_j(\x)\,d\x$ satisfy the relations in Remark \ref{volumefracs}. However, it is not possible to check these relations experimentally from the area fractions of the variants on a single slice of the grain as in Fig. 3. Furthermore, by Corollary \ref{rigidv},  we have that $\bar\nu_\x=\bzero$ for a.e. $\x\in\om$, so that
\be
\label{averagegradient}
\sum_{j=1}^{12} \lambda_j(\x)\,{\ab_j\otimes\n_j}=\bzero\;\text{  for a.e. }\x\in\om
\ee
or equivalently
\be
\label{averagegradient1}
\sum_{j=1}^{12}\lambda_j(\x)\,(\V_j-\1)=\bzero\;\text{  for a.e. }\x\in\om.
\ee
In particular, by Fubini's theorem, for any unit vector $\N$ we have that, for almost all $t\in\R$, 
\be
\label{averagegradient2D}
\sum_{j=1}^{12}\lambda_j(\x)\,(\V_j-\1)=\bzero
\ee
for a.e. $\x$ in the two-dimensional slice $\om_t:=\{\z\in\om:\z\cdot \N=t\}$. Hence, taking $\om_t$ to be the specimen surface in Fig. \ref{fig3} with $\N$ parallel to [100 89 19] we expect that for any 2D ball $B_2(\x_0,r)\subset \om_t$, with small but not too small radius $r>0$, and in which the $i^{\rm th}$ variant $\V_i$ has volume fraction $\rho_i(\x_0,r)$, the average deformation gradient
\be
\label{avgrad}
\av_{B_2(\x_0,r)}D\y(\x)\,d\x\approx\sum_{i=1}^{12}\rho_i(\x_0,r)\V_i \sim\1,
\ee
and this prediction is confirmed by the calculated averages in Fig \ref{fig5}.

In order to consider more general boundary conditions consistent with a zero-energy microstructure we need to know $K^{\rm qc}$. Indeed by \cite[Theorem  6.1]{kinderlehrerpedregal91} and the fact (see e.g. \cite[Theorem 4.10]{muller99}) that $K^{\rm qc}$ is the set of centres of mass $\bar\nu$ of homogeneous gradient Young measures $\nu$ with $\supp\nu\subset K$,  a map $\y\in W^{1,\infty}(\om;\R^3)$ satisfies $D\y(\x)\in K^{\rm qc}$ for a.e. $\x\in\om$ iff there is a gradient Young measure $(\nu_\x)_{\x\in\om}$ with $\supp\nu_\x\subset K$ and underlying macroscopic deformation $\y$.  However, as explained in Section \ref{analysis} we are unfortunately not able to determine $K^{\rm qc}$.

\subsection{The case $\alpha\beta\neq 1$.}
If $\alpha\beta\neq 1$ then there is no gradient Young measure $(\mu_\x)_{\x\in\om}$ with underlying macroscopic deformation $\y$  satisfying \eqref{GYM} and \eqref{bc}. Indeed we would then have that $D\vc y(\x)\in H^{\rm qc}\subset H^{\rm pc}$ for a.e. $\x\in\om$, so that by Corollary \ref{rigidv}  $((\alpha\beta)^\frac{1}{3}-1)\1=\vc a\otimes\vc n$, which is impossible.  Thus any zero-energy microstructure must deform the grain boundary other than by a uniform dilatation.

\subsection{Stressed microstructures and dynamics}
\label{stressed}
The experimental observations of exact interfaces between different martensitic variants in the {\Tn}  microstructure imply that the deformation gradient takes some values with positive energy and nonzero stress, despite its overall closeness to $H$.
This is consistent  with   Theorem \ref{MS THM}, which, as explained after the theorem, implies  in the case $\alpha\beta=1$ that for any $\ep>0$ there is a deformation $\y_\ep$ satisfying the boundary condition $\vc y_\ep(\vc x) =\vc x$  on $\partial\Omega$, as well as small perturbations of this boundary condition, with $\dist (D\y_\ep(\x),H)<\ep$ for a.e. $\x\in\om$.
 
 That the microstructure is slightly stressed is also consistent with  the austenite start temperature $A_s$ being less than $\theta_c$. If the martensitic microstructure were at zero energy then there would be no reason to transform back to the higher energy austenite at a temperature less than $\theta_c$.  However a complete analysis of this would require an understanding of the competing roles of interfacial energy, ignored in our model, since the reverse transformation can be expected to reduce the area of martensite-martensite interfaces, thus providing a driving force for the transformation, while introducing austenite-martensite interfaces that act as an additional energetic barrier.

The martensitic microstructure is formed dynamically by a process that is not directly experimentally observed in 3D. However some insight can be gained both from the observed microstructure, as described in Section \ref{expresults}, and from  {\it in situ} optical microscopy with a high-speed camera of the formation of microstructure at a free surface in a similar alloy \cite{inamura2017}. These observations suggest that the martensite first forms colonies consisting of three variants with very small incompatibility. Once a colony forms, its plates partition the surrounding space into smaller regions. Because each variant can serve as a bridge to generate another colony, the process continues in a chained manner, with colonies successively nucleating and growing from one another. Inside the increasingly fine regions created by this partitioning, further colonies develop in a nested way.

  From a theoretical perspective, a  frame-indifferent dynamical model allowing for sharp interfaces that is known to be well posed is not currently available, but some insight can be gained from approximate dynamical models. 

One such approximate model  (see \cite{DellaPorta2019}) is based on  the `moving mask' assumption, motivated by dynamic observations by Song et al \cite{songetal2013} of thermal cycling in the alloy ${\rm Zn}_{45}{\rm A}_{30}{\rm Cu}_{25}$, whereby the martensitic microstructure remains unchanged once it is formed, so that it appears to be unveiled as if by a moving mask. In \cite{DellaPorta2019a} this  model is  framed within the context of   nonlinear elastodynamics.  It is shown in \cite{DellaPorta2019} that the moving mask assumption  implies that at each point of the martensitic microstructure the deformation gradient must have the form 
\be
\label{movingmask}
D\y(\x)=\1+\ab(\x)\otimes\n(\x).
\ee
Thus in considering deformations with $D\y(\x)\in H$ and gradient Young measures $(\nu_{\x})_{\x\in\om}$ with $\supp\nu_\x\in H$ for a.e. $\x\in\om$ we have implicitly used the moving mask assumption, this approximation being experimentally supported by the observed closeness of the deformation gradient to $H$.  

Another approximate model is that introduced in \cite{p36} (see also \cite{DellaPortaetal21}), which attempts to understand the sequential development of martensitic microstructure via random nucleation of infinitely thin martensitic plates with specific orientations. Combining this with the moving mask approximation leads to a picture whereby the martensitic microstructure evolves by the nucleation at random sites in the remaining austenite of plates having close to one of the twelve deformation gradients in $H$, each taken with an equal probability, until the transformation is complete.  However, the nucleation process of the martensite in ${\rm Zn}_{45}{\rm A}_{30}{\rm Cu}_{25}$, and thus presumably also in \Tn, is more complicated than that assumed in the probabilistic model, in that the martensite nucleates in pairs formed from the most compatible variants.

 In \cite{cesanahambly2022} the probabilistic model in 2D with only horizontal and vertical orientations, and the corresponding 3D model with three perpendicular orientations, are analyzed as  branching random walks, and asymptotic power laws for the length (area in 3D) distribution of plates are obtained. This is interesting in view of the power law distribution of the lengths of plates observed in \Tn, for which the number $N(L)$ of plates of length $L$ is found to be proportional to $L^{-\theta}$, where the exponent $\theta\approx 2.6$ over a wide range of grain sizes and for different orientations of the 2D grain surface with respect to the cubic axes. This exponent is not currently predicted by any theory.\\ \\

\noindent{\bf Appendix: Experimental procedures}

Ti, Nb and Al with a purity of 99.99\% or higher were used as starting materials. Ingots with a nominal composition of 76Ti-22Nb-2Al (mol\%) were fabricated in an Ar-1\%H2 atmosphere by non-consumable electrode arc-melting. To prevent segregation, the melted ingots were turned over and melted again a total of six times. The mass change of the ingot before and after melting was less than 0.04 wt\%, so that an alloy with the target composition was considered to have been obtained. The ingots were vacuum-sealed in quartz tubes, homogenized ($1273$ K, 7.2 ks) and then quenched in iced water. Cubes of approximately 10 mm per side were cut out by electrical discharge machining and cold-rolled from three directions to 90\% of cumulative strain in order to eliminate the solidified structure. Solution treatment in vacuum (1173 K, 1.8 ks, quenched in iced-water) was then carried out. The martensite start ($M_s$), finish ($M_f$), austenite start ($A_s$) and finish ($A_f$) temperatures of the specimen prepared in this procedure were 408 K, 314 K, 375 K and 460 K, respectively.

The lattice parameters were determined by $\theta-2\theta $ X-ray diffraction (XRD) measurements. A Philips X'pert-Pro Galaxy system (CuKa) with Anton Paar TTK450 temperature control unit was used. The systematic error was corrected using an internal standard method with Cu. The specimen surfaces were electropolished (${\rm HClO}_4$ : ${\rm CH}_3{\rm(CH_2})_2){\rm CH}_2{\rm OH} : {\rm CH}_3{\rm OH} = 1 : 6 : 10$, $233\sim 243$ K, 30 V) and subjected to X-ray diffraction.  Diffraction patterns were taken at $10$ K intervals below $M_s$,  the closest measurement temperature to $M_s$ being  $\theta=$403 K, in the range of $20^\circ<2\theta < 120^\circ$. The lattice parameters $\alpha,\beta,\gamma$ were determined by the least-squares method and the measured values are given in Table \ref{lattice}.
\begin{table}[h]\hspace{1in}
\begin{tabular}{|c|c|c|c|}
\hline
Temperature K&$\alpha$     & $\beta$        & $\gamma$              \\ \hline 403&0.9699&1.0305&1.0003\\ \hline 393&0.9705&1.0300&1.0001\\ \hline
383&0.9700&1.0313&1.0012\\ \hline 373&0.9727&1.0338&1.0025\\ \hline
363&0.9711&1.0343&1.0030\\ \hline\end{tabular}
\caption{Measured values of $\alpha,\beta,\gamma$, accurate to $\pm 0.003$.}
\label{lattice}
\end{table}
The  lattice parameter values at $M_s$ and $\theta_c$  given in Section \ref{model} were linearly extrapolated from those in Table \ref{lattice}.

A field-emission gun-type scanning electron microscope (FE-SEM, Hitachi SU-5000) with electron backscatter diffraction (EBSD, TSL OIM DVC5) detector and back scattering electron (BSE) detector was used. The surface of the specimen for the microscopy was finished by electropolishing with the same solution and conditions as the XRD specimens.  
The 12 variants were identified by the orientation imaging microscopy (OIM) software. The orientation of the prior austenite phase was reproduced from the average position of the (100), (010), (001) poles of all the variants.  Martensite variants were identified using the theoretical orientations of the 12 variants calculated from the reproduced orientation of the parent phase.\\
\section*{AI tools and computational resources disclosure} The paper was written entirely by the authors, but checked for consistency with Gemini 3. The computation of the average deformation gradients in Fig. \ref{fig5} was aided by Chat GPT.  Numerical computations related to Table \ref{thetatable} were carried out using MATLAB.



\section* {Acknowledgements}  The authors are grateful to Angkana R\"{u}land for encouraging us to write this paper and to Bernd Kirchheim for many useful discussions. JMB was supported by EPSRC grant EP/V00204X on  the {\it Mathematical theory of polycrystalline materials}, by the Hong Kong Institute for Advanced Study, and by the School of Mathematics at the Institute for Advanced Study in Princeton. FDP contributed to this research mostly during a postdoc at the Max Planck Institute for Mathematics in the Sciences in Leipzig. TI was supported by JSPS KAKENHI (26H02196), JST A-STEP (JPMJTR24T6), JST CREST (JPMJCR26D1), and the Light Metal Educational Foundation (2024-C2). He would also like to thank Daisuke Osanai for his contribution to the microscopy observations.
 \bibliographystyle{plain}
 
\bibliography{balljourn,ballconfproc,gen2,ballprep}

\end{document}